\documentclass[preprint]{aastex}
\usepackage{url}
\shorttitle{EB Science with Extrasolar Planet Surveys}
\shortauthors{Fleming et al.}
\begin{document}

\title{Eclipsing Binary Science Via the Merging of Transit and Doppler Exoplanet Survey Data - A Case Study With the MARVELS Pilot Project and SuperWASP}

\author{Scott W. Fleming\altaffilmark{1}, Pierre F. L. Maxted\altaffilmark{2}, Leslie Hebb\altaffilmark{3}, Keivan G. Stassun\altaffilmark{3,4}, Jian Ge\altaffilmark{1}, Phillip A. Cargile\altaffilmark{3}, Luan Ghezzi\altaffilmark{5,17}, Nathan M. De Lee\altaffilmark{1}, John Wisniewski\altaffilmark{6}, Bruce Gary\altaffilmark{3}, G.~F. Porto de Mello\altaffilmark{7,17}, Leticia Ferreira\altaffilmark{7,17}, Bo Zhao\altaffilmark{1}, David R. Anderson\altaffilmark{2}, Xiaoke Wan\altaffilmark{1}, Coel Hellier\altaffilmark{2}, Pengcheng Guo\altaffilmark{1}, Richard G. West\altaffilmark{8}, Suvrath Mahadevan\altaffilmark{9,10}, Don Pollacco\altaffilmark{11}, Brian Lee\altaffilmark{1}, Andrew Collier Cameron\altaffilmark{12}, Julian C. van Eyken\altaffilmark{1,13}, Ian Skillen\altaffilmark{14}, Justin R. Crepp\altaffilmark{1,15}, Duy Cuong Nguyen\altaffilmark{1}, Stephen R. Kane\altaffilmark{1,13}, Martin Paegert\altaffilmark{3}, Luiz Nicolaci da Costa\altaffilmark{5,17}, Marcio A.~G. Maia\altaffilmark{5,17}, Basilio X. Santiago\altaffilmark{16,17}}
\email{scfleming@astro.ufl.edu}
\altaffiltext{1}{Department of Astronomy, University of Florida, 211 Bryant Space Science Center, Gainesville, FL 32611, USA}
\altaffiltext{2}{Astrophysics Group, Keele University, Staffordshire, ST5 5BG, UK}
\altaffiltext{3}{Department of Physics and Astronomy, Vanderbilt University, Nashville, TN 37235, USA}
\altaffiltext{4}{Department of Physics, Fisk University, 1000 17th Ave. N., Nashville, TN 37208, USA}
\altaffiltext{5}{Observat\'{o}rio Nacional, Rua General Jos\'{e} Cristino, 77, 20921-400 S\~{a}o Crist\'{o}v\~{a}o, Rio de Janeiro, RJ, Brazil}
\altaffiltext{6}{Department of Astronomy, University of Washington, P.O. Box 351580, Seattle, WA 98195, USA}
\altaffiltext{7}{Universidade Federal do Rio de Janeiro, Observat\'{o}rio do Valongo, Ladeira do Pedro Ant\^{o}nio, 43, CEP: 20080-090, Rio de Janeiro, RJ, Brazil}
\altaffiltext{8}{Department of Physics and Astronomy, University of Leicester, Leicester, LE1 7RH, UK}
\altaffiltext{9}{Department of Astronomy and Astrophysics, The Pennsylvania State University, 525 Davey Laboratory, University Park, PA 16802, USA.}
\altaffiltext{10}{Center for Exoplanets and Habitable Worlds, The Pennsylvania State University, University Park, PA 16802, USA.}
\altaffiltext{11}{Astrophysics Research Centre, School of Mathematics \& Physics, Queen's University, University Road, Belfast, BT7 1NN, UK}
\altaffiltext{12}{SUPA, School of Physics \& Astronomy, University of St Andrews, North Haugh, KY16 9SS, St Andrews, Fife, Scotland, UK}
\altaffiltext{13}{NASA Exoplanet Science Institute, Caltech, MS 100-22, 770 South Wilson Avenue, Pasadena, CA 91125, USA}
\altaffiltext{14}{Isaac Newton Group of Telescopes, Apartado de correos 321, E-38700 Santa Cruz de la Palma, Canary Islands, Spain}
\altaffiltext{15}{Department of Astronomy, California Institute of Technology, 1200 E. California Blvd., Pasadena, CA 91125, USA}
\altaffiltext{16}{Instituto de F\'{i}sica, UFRGS, Caixa Postal 15051, Porto Alegre, RS - 91501-970, Brazil}
\altaffiltext{17}{Laborat\'{o}rio Interinstitucional de e-Astronomia, - LIneA, Rua Gal. Jos\'{e} Cristino 77, Rio de Janeiro, RJ - 20921-400, Brazil}

\begin{abstract}
Exoplanet transit and Doppler surveys discover many binary stars during their operation that can be used to conduct a variety of ancillary science.  Specifically, eclipsing binary stars can be used to study the stellar mass-radius relationship and to test predictions of theoretical stellar evolution models.  By cross-referencing 24 binary stars found in the MARVELS Pilot Project with SuperWASP photometry, we find two new eclipsing binaries, TYC 0272-00458-1 and TYC 1422-01328-1, which we use as case studies to develop a general approach to eclipsing binaries in survey data.  TYC 0272-00458-1 is a single-lined spectroscopic binary for which we calculate a mass of the secondary and radii for both components using reasonable constraints on the primary mass through several different techniques.  For a primary mass of $M_1 = 0.92 \pm 0.1 ~ M_{\odot}$, we find $M_2 = 0.610 \pm 0.036 ~ M_{\odot}$,  $R_1 = 0.932 \pm 0.076 ~ R_{\odot}$ and $R_2 = 0.559 \pm 0.102 ~ R_{\odot}$, and find that both stars have masses and radii consistent with model predictions.  TYC 1422-01328-1 is a triple-component system for which we can directly measure the masses and radii of the eclipsing pair.  We find that the eclipsing pair consists of an evolved primary star ($M_1 = 1.163 \pm 0.034 ~ M_{\odot}$, $R_1 = 2.063 \pm 0.058 ~ R_{\odot}$) and a G-type dwarf secondary ($M_2 = 0.905 \pm 0.067 ~ M_{\odot}$, $R_2 = 0.887 \pm  0.037~ R_{\odot}$).  We provide the framework necessary to apply this analysis to much larger datasets.
\end{abstract}

\section{Introduction}
\label{intro}
Photometric and Doppler radial velocity (RV) surveys have generated comprehensive data sets in their quest for extrasolar planets.  These surveys have been successful in discovering several hundred planets\footnote{\url{http://exoplanet.eu/}} and furthering our understanding of how planets form, their dynamically evolution, and their distribution of physical properties.  These data sets also provide a wealth of ancillary science studies involving variable stars, binary stellar systems and brown dwarf companions.  The amplitudes of the photometric or Doppler variability from such objects are often several orders of magnitude larger than extrasolar planet signals, making them comparatively simpler to identify and characterize.

Cross-referencing stars in both photometric and Doppler surveys can minimize the amount of follow-up data required to conduct these ancillary projects \citep[e.g.,][]{kan2009}.  For binary systems, there are several advantages in combining data from both types of surveys.  One disadvantage of using spectroscopic binaries (SBs) as a source of detached eclipsing binary (EB) candidates is that the line-of-sight inclination of the orbital plane is unconstrained.  The a priori probability that a given SB eclipses can be estimated as $\left(R_{1}+R_{2}\right) / a$, where $R_{1}$ and $R_{2}$ are the radii of the primary and secondary, respectively, and $a$ is the orbital semimajor axis.  Typical a priori values range from a few percent up to $\sim 25 $\%, therefore conducting photometric follow-up results in null detections a majority of the time.

In addition, the predicted ephemerides for eclipses based on modeling the Doppler RV data depend sensitively on the orbital phase coverage, the precise values of the eccentricity and argument of periastron, and the amount of time elapsed since the last measurement.  Conversely, archival data from photometric surveys consist of tens of thousands of epochs over several years, resulting in nearly continuous phase coverage for short orbital periods.  Given the orbital period from the spectroscopic orbit, it is more efficient to check archival photometry from transit surveys than to attempt follow-up photometric observations based on the transit ephemeris from the radial velocity data.

Attacking the problem from the other direction results in a different set of drawbacks.  The advantage of starting with an EB detected from a photometric survey is that the favorable geometry for an eclipse is already known to exist.  However, obtaining high-precision, time-series RV follow-up is a resource-intensive enterprise.  Such observations are conducted using heavily-subscribed instruments (generally Echelle spectrographs) on moderately large telescopes (usually larger than two meters in diameter) for which access to observing time is quite competitive.  Obtaining the best orbital solution nominally requires a minimum of seven epochs that sample the orbital phase well, requiring many nights of follow-up observations.

Doppler RV surveys, on the other hand, already have the necessary number of epochs and sampling rate because an identical observing strategy is required to detect planets.  The observing cadences are selected with great care in order to provide synoptic coverage of all possible exoplanet orbital periods and phases.  Some surveys cease observations if a star exhibits Doppler variation of many $\rm{km ~ s^{-1}}$ over short timescales, indicative of an SB, to maximize planet yield.  Other surveys, such as the multi-object MARVELS survey \citep{ge2009}, continue to observe the SB to fully characterize the orbit.  For the specific case of MARVELS, the multiplicity of a given target is usually not known prior to the start of observations.  Continuing to observe SBs to fully characterize their orbit works particularly well in multi-object surveys, where keeping a few SBs per field maximizes science results while not significantly decreasing the capability of detecting extrasolar planets.  

Another disadvantage of EBs that only have photometric survey data is that the masses of the two components can only be estimated.  Although the mass of the primary can be estimated using colors \citep[e.g., the infrared flux method,][]{cas2010}, deriving the mass of the secondary usually requires an observation of the reflex motion of the primary.  Given the large number of EBs present in transit surveys, the ability to select a subset that have the precise masses of interest for the project is more efficient.

\section{The Mass-Radius Relationship of K and M Dwarfs}
\label{massradkm}
One example of binary science that benefits greatly from the merging of Doppler and photometric survey data is the study of the stellar mass-radius relationship.  Observations of EBs have been in excellent agreement with theoretical models with the exception of K and M dwarfs \citep[$M<0.8 ~ M_{\odot}$,][]{lop2007,tor2010}.  Stars in this mass range have partially convective outer atmospheres, and can have radii that are 10-20\% larger than models predict.  Stars above this mass range have primarily radiative atmospheres and agree well with current models \citep[e.g.,][ and additional references in \citet{tor2010}]{pop1997,imb2002,tor2008,mei2009}.

One possible cause of this inflation is the stronger magnetic fields for K and M dwarfs in short-period binaries as compared to isolated stars of the same mass \citep{cha2007}.  Tidal synchronization increases the rotational period of the stars to match the orbital period, and the stronger magnetic dynamo acts to suppress convection in the outer layers of the star.  This results in an increase in cool starspots, reducing the effective temperature, and therefore the radius of the star expands to maintain the luminosity of the star.  Recent theoretical work has shown that starspots predominantly distributed at polar latitudes, and/or suppression of convective transport in the outer layers of the atmosphere, are both effects caused by magnetic fields that can reproduce the observations \citep{mor2010}.

In order to place constraints on the interaction of magnetic fields and the stellar radius, a large sample of EBs with masses and radii measured to precisions of a few percent is required.  Specifically, the sample should have at least one component in each EB that is $0.3<M<0.8 ~ M_{\odot}$ and should have a wide range of orbital periods in order to sample a range of magnetic field strengths.  However, despite several decades of surveys for photometric variability, there remains a paucity of such EB systems with precisely determined masses and radii.  For example, there were only 13 EB systems in the compilation of \citet{lop2007} that fit the mass criterion above.  Combining photometric and Doppler survey data can help increase the sample size by factors of several over the next few years, particularly at longer orbital periods ($P > 5~$days).  Ground-based surveys for exoplanets identify many SBs and EBs that are, by design, bright stars (V $<$ 12).  This translates to higher signal-to-noise, requiring less telescope time for follow-up, and therefore a larger number of these binaries can be measured with high precision.  The Kepler mission is uncovering thousands of EBs with exquisite photometric precision \citep{cou2010,prs2010}, but a majority of those binaries are several magnitudes fainter than EBs found from ground-based exoplanet surveys, resulting in resource-intensive spectroscopic follow-up.  In addition, any such ground-based observations are in direct competition with programs dedicated to exoplanet follow-up.

Although several groups are working on studying the mass-radius relationship of double-lined spectroscopic binaries (SB2s) from ground-based surveys \citep[e.g.,][]{dev2008}, less attention is placed on binaries consisting of F/G primaries and K/M secondaries \citep[but see][]{fer2009}.  These binaries are generally brighter than K/M pairs, and avoid any biases introduced by studying binaries with mass ratios close to unity.  The difficulty with studying binaries that have small flux ratios ($F_2 ~ F_1^{-1}$) is that the mass of the fainter component usually cannot be measured using the survey data.  In order to measure the Doppler motion of the fainter companion, large-aperture telescopes equipped with high resolution spectrographs must be used in conjunction with two-dimensional correlation techniques \citep[TODCOR,][]{zuc1994,zuc2003}.  Only a minimum number of epochs per target using these large telescopes are required, because other parameters such as the orbital period, semiamplitude of the primary, and epoch of transit are known from the survey data.  Observations can be taken near the RV extrema to measure the semiamplitude of the fainter component and thereby obtain both dynamical masses.

As a case study, we present initial results from a cross-referencing of 24 SBs with $P < 20~$days from the MARVELS Pilot Project (hereafter MPP) with photometry from the SuperWASP transit survey \citep{pol2006}.  This cross-referencing was performed after the RV observations from the MPP were completed, and the SuperWASP lightcurves were visually examined to identify which of the SBs were eclipsing.  Out of those 24 SBs, five did not have a sufficient number of photometric observations and two are blended with other stars.  Out of the remaining 17 SBs we have found two EBs: TYC 0272-00458-1 and TYC 1422-01328-1, which we refer to as TYC 272 and TYC 1422 throughout the rest of the paper.  TYC 272 is a single-lined spectroscopic binary (SB1), but we constrain the mass of the primary using several different techniques to estimate the properties of the host star.  We then calculate the mass of the secondary, as well as the radii of the primary and secondary, from the survey data.  TYC 1422 is found to be a triple system with all three components visible in high-resolution spectra.  We note that although this work analyzed the photometry and spectroscopic data separately, many software packages exist to simultaneously model all observational data to derive the best-fit binary solutions, \citep[e.g.,][]{prs2005}.  We demonstrate the success of cross-referencing Doppler and transit survey data and lay the foundations for more comprehensive studies using significantly larger data sets.

\section{Survey Data}
\label{obsdata}

\subsection{MPP Doppler Observations}
\label{mppdata}
The MPP was a trial survey conducted in 2007 that used the Keck ET instrument \citep{gespie2006} on the Sloan Digital Sky Survey (SDSS) 2.5m telescope \citep{gun2006} at the Apache Point Observatory.  Although the primary objective was to serve as a test study for the MARVELS extrasolar planet survey, the MPP obtained between 5 and 38 RV measurements of 708 stars over 1-5 month baselines.  The 59-object, fiber-fed instrument observed a total of twelve fields, and successfully demonstrated the ability to conduct a multi-object, DFDI \citep[dispersed fixed-delay interferometry,][]{gee2002,ge2002,ers2002,ers2003} survey for extrasolar planets.  The spectral resolution of the Keck ET was $R \sim 5100$ and had a wavelength range of $495 < \lambda < 585 ~$ nm.  Despite its relatively small scale, the MPP was also able to demonstrate the ancillary science that can be done with such a survey, including the discovery of a short-period, brown dwarf candidate with minimum mass $m_{\rm{min}} = 64.3 ~ M_{\rm{Jupiter}}$ \citep{fle2010}.

\subsection{SuperWASP Photometry}
\label{swdata}
The WASP instruments provide flux measurements for millions of stars using wide-angle images of the night sky over a band pass of 400-700 nm defined by a broad-band filter.  The survey has produced more than 35 transiting exoplanets to-date, orbiting stars of A through K spectral types and with orbital periods $0.78 < P < 8.1$ days \citep[e.g.,][]{col2007a,col2010,heb2010,que2010}, in addition to studies of variable stars \citep{nor2007,ant2010,wil2010}.  Eight cameras on each instrument provide images covering approximately 7.8x7.8 degrees using Canon 200mm f/1.8 camera lenses and e2v 2048x2048 CCDs. Synthetic aperture photometry using an aperture radius of 49 arcseconds at the position of cataloged stars is performed on the images \citep{pol2006}. We extracted 1003 observations from the WASP archive for TYC 272 from 10 nights with good quality data covering either the primary or secondary eclipse. For TYC 1422, 649 data points from 12 nights of good quality data were used.

\section{Follow-up Observations}
\label{followup}

\subsection{Hereford Arizona Observatory Absolute Photometry}
\label{haodata}
We used the Hereford Arizona Observatory (HAO), a private facility in southern Arizona (observatory code G95 in the IAU Minor Planet Center) to measure multi-band, absolute photometry of TYC 272 and TYC 1422.  The facility has been used to measure transiting exoplanets with high photometric precision\footnote{\url{http://brucegary.net/book_EOA/x.htm}}, and has recently been used to observe EBs and variable star lightcurves with similar levels of precision.  HAO employs an 11-inch Celestron Schmidt-Cassegrain (model CPC 1100) telescope fork-mounted on an equatorial wedge and located in a dome.  The 11-inch telescope's CCD is an SBIG ST-8XE CCD with a KAF 1602E detector.  HAO also includes a 14-inch Meade Schmidt-Cassegrain (model LX200GPS) telescope, fork-mounted on an equatorial wedge located in a dome.  That telescope's CCD is an SBIG ST-10XME with a KAF-3200ME detector.  A 10-position filter wheel accommodates SDSS and Johnson/Cousins filter sets.

The stars were observed with Sloan $g^\prime,r^\prime,i^\prime$ filters with the 11-inch telescope.  To derive the target star's magnitude in band $x$ we use a generic photometry equation:
\begin{equation}
M_x = M_{x_{0}} - 2.5 \log \left( \frac{F_x}{t} \right) - K^{\prime}_{x} z + S_{x} C
\label{photeq}
\end{equation}
where $M_x$ is the desired magnitude for observed band $x$, $M_{x_0}$ is an instrumental zero-shift for band $x$, $F_x$ is the measured intensity in the observed band, $t$ is the integration time, $K^{\prime}$ is the zenith extinction for the observed band, $z$ is the airmass, $S_x$ is the star color sensitivity for the observed band and $C$ is a star color defined using two bands (e.g., $(B-V)$ or $(g^{\prime}-r^{\prime})$).  For the Johnson-Kron-Cousins bands, standard stars are taken from the list published by \citet{lan2007} and \citep{lan2009}. For the SDSS bands, standard stars are taken from the list published by \citet{smi2002}.  $B$ and $V$ magnitudes are determined using the conversion equations given in \citet{smi2002} that convert $g^{\prime}$, $r^{\prime}$ and $i^{\prime}$ magnitudes to $B$, $V$ and $R_c$, while $I_c$ is determined using the $i^{\prime}$ fluxes.

\subsection{SMARTS Echelle Spectroscopy}
\label{smartsdata}
We obtained spectra from the SMARTS 1.5m telescope at CTIO using the R $\sim$ 42000 Echelle spectrograph that covers $402 < \lambda < 730 ~$ nm.  These observations were used to resolve the components of TYC 1422 and get individual RV measurements for each star in the system.  Seven epochs were observed using ten minute integrations between February and April 2010 with typical signal-to-noise ratios of $\sim$ 20 per resolution element at $\lambda = 600$ nm.

The spectra were processed in the standard way for cross-dispersed Echelle spectra, using a pipeline written specifically for spectra taken with the Echelle spectrograph on the CTIO-1.5m.  The routine processes the data using biases and quartz lamp observations taken at the beginning of the night, median combines three individual images while performing cosmic ray rejection, extracts the individual orders from the combined image and performs the wavelength solution on each order using a ThAr arc lamp taken either before or after each set of science exposures.

A total of 22 orders are used each night to derive radial velocities via cross-correlation with a standard template.  We use the bright G0 star HD 108510 as the heliocentric RV standard star.  Each spectral order is cross-correlated separately, then an iterative 3-$\sigma$ clipping is performed prior to performing a weighted average to obtain a final RV measurement for each night.  The components are identified each night via the peak and width of each feature in the CCF.  The typical RV precisions ranged from 0.3 to 1.7 $\rm{km~s^{-1}}$ for the various components.

\subsection{APO Echelle Spectroscopy}
\label{arcesdata}
We obtained ARCES \citep{wan2003} spectroscopy from the APO 3.5m telescope to check for additional spectra from unresolved companions.  ARCES is an Echelle spectrograph with spectral resolution $R \sim$31500 that covers the entire wavelength range from $320 < \lambda < 1000 ~$ nm. on a single 2048x2048 SITe CCD.  Three exposures for a total integration time of 18 minutes were obtained for TYC 1422 on 2009 March 23, while two 20-minute exposures were taken of TYC 272 on 2010 June 19.  The spectra were reduced using an IRAF script.  Spectra are corrected for bias and dark subtraction, cosmic rays and bad pixels.  Flatfielding is done using a combination of two different sets of quartz lamp exposures; one with a blue filter in place and another set without the filter.  These two different sets of flats allow for optimal extraction of both the blue and red orders.  A ThAr lamp is used for wavelength calibration.  The co-added spectrum for TYC 1422 had a signal-to-noise of $\sim 75$ per resolution element at 600 nm, while TYC 272 had a signal-to-noise of $\sim 105$ per resolution element.

\section{Case Studies}
\subsection{TYC 272}
\subsubsection{Orbital Parameters}
\label{0272orbit}
We first identified the binaries spectroscopically using the MPP RV data.  Following \citet{fle2010}, the RV uncertainties for each star are scaled by a ``quality factor'' that is a multiplicative scaling that modifies the formal RV uncertainties to better account for systematics defined as:
\begin{equation}
Q = \frac{\rm{RMS}\left(X - \langle {X} \rangle\right)}{\rm{MEDIAN}\left(\sigma_{X}\right)},
\label{eqn:qf}
\end{equation}
where $X$ represents the RV measurements and RMS is the root-mean-square residual.  The quality factor is derived from the RMS of the RVs for the other 58 objects in the same field-of-view as the target, under the assumption that a majority of the objects per field should not have significant RV variability.  The field containing TYC 272 has a $Q$ = 3.44, suggesting the formal uncertainties are underestimated.  Table \ref{0272rvs} contains the times of observations, RVs, and both the formal and scaled RV uncertainties for TYC 272.  A total of 14 epochs were obtained with an average, scaled RV uncertainty of 0.345 $\rm km ~ s^{-1}$ for TYC 272.

\begin{figure}
\plotone{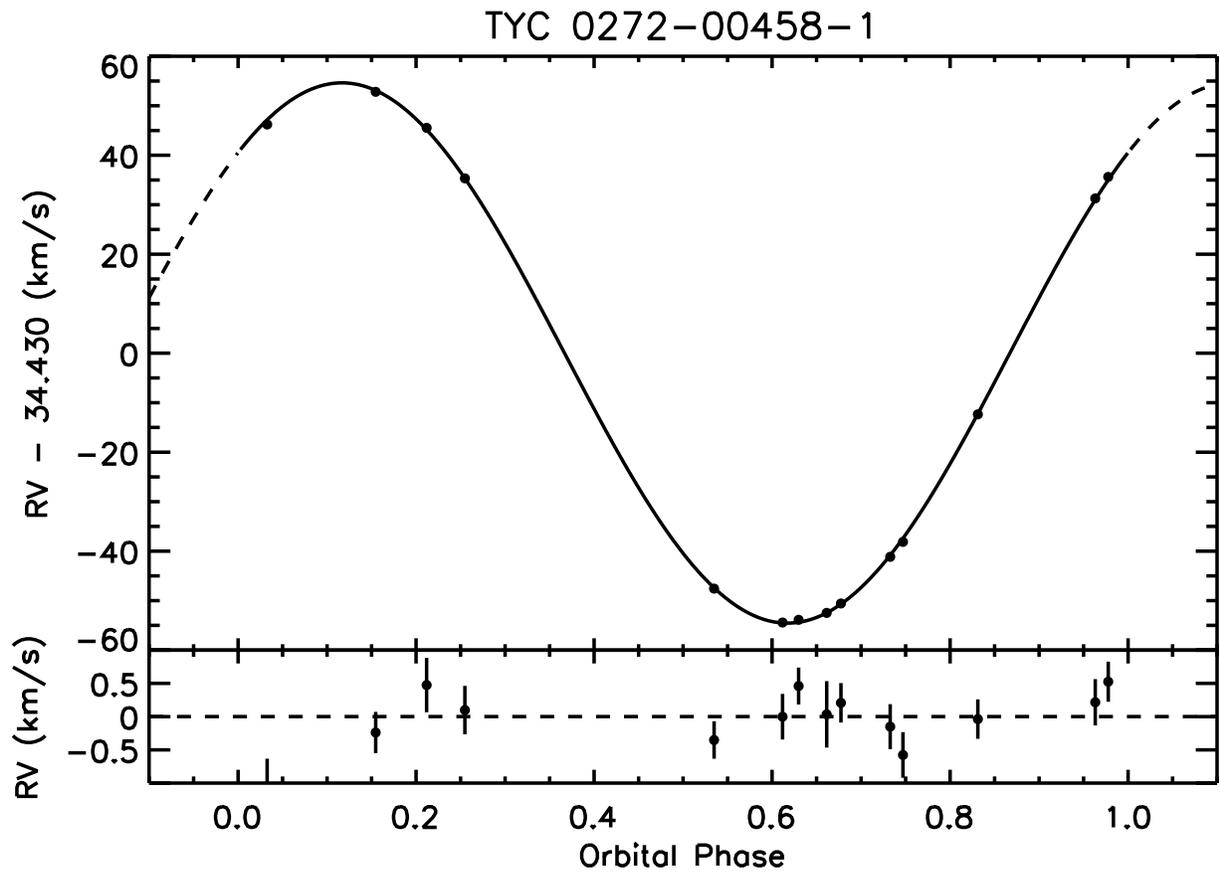}
\caption{Best-fit orbital solution for TYC 272 from the MPP data.  The RV uncertainties have been scaled by a factor of 3.44 to account for additional systematic errors.\label{tyc0272folded}}
\end{figure}

TYC 272 was fit using the RVSIM software program \citep{kan2007}.  Fig.\ \ref{tyc0272folded} shows the best-fit orbital solution to the MPP RV data.  The $\chi^2 / \rm{dof}$, after scaling by $Q$, is 2.62 with eight degrees of freedom.  Because it is a short-period binary, there is additional RV jitter expected due to chromospheric activity, however, chromospheric effects are typically on the order of tens of $\rm{m~s^{-1}}$ \citep{san2000}.  The eccentricity is consistent with a circularized orbit, as expected for the best-fit orbital period $P = 5.7282$ days \citep{maz2008}.  The semi-amplitude $K \sim 54.6 ~ \rm{km ~ s^{-1}}$ corresponds to a minimum mass of the secondary $m_{\rm{min}} \sim 0.64 ~ M_{\odot}$ assuming the mass of the primary is $M = 1 ~ M_{\odot}$, making this EB an excellent system to further study in the context of the mass-radius relationship.  The best-fit orbital properties are listed in Table \ref{tyc0272_dynprop}.

\subsubsection{Stellar Properties}
\label{0272starprops}
We utilize a spectral synthesis technique as part of the process to estimate basic stellar parameters such as $T_{\rm eff}$, $\log{(g)}$ and $\rm{[Fe/H]}$ for the primary.  In \citet{val2005}, they outline a general procedure for deriving stellar properties using Spectroscopy Made Easy \citep[SME;][]{val1996}.  We follow the exact methodology described in \citet{val2005}, but include some changes that allow for a more in-depth search of $\chi^2$ space.  Specifically, we use the same spectral regions, line lists and continuum regions as in \citet{val2005}, but do not allow any of the individual abundances to be free parameters.

Our SME spectral analysis pipeline is divided into two steps, because the spectral synthesis parameters are highly correlated, and therefore a range of initial guesses is required to adequately explore the $\chi^2$ space.  First, a forward modeling procedure is used to calculate $\chi^2$ values for model spectra over a large grid of parameter space. This allows us to focus our analysis on the region of parameter space with the most likely, global, best-fit stellar parameter solution. Second, we use the SME Levenberg-Marquardt minimization algorithm as was used in \citet{val2005} to fit observed spectra with synthetic spectra for 100 different initial input parameters. We select these different combinations of input values from the 100 grid points with the lowest chi-square values determined from the first step of our analysis pipeline. Accessing Vanderbilt University’s ACCRE parallel computer facility allows us to efficiently explore the dependence of these initial parameters on our output parameters, as well as providing a good measure of the internal precision of SME’s minimization procedure.

Using our implementation of SME on the ARCES spectrum of TYC 272, we derive best-fit values of $T_{\rm eff} = 5905 \pm 177$ K, $\log{(g)} = 4.83 \pm 0.19$, metallicity $\rm{[Fe/H]} = -0.28 \pm 0.12$.  The value for $\log{(g)}$ is larger than expected for stars of this temperature, but we note that $\log{(g)}$ is a very challenging parameter to measure spectroscopically, even at high spectral resolution, and is particularly challenging at the ARCES spectral resolution of $\sim$ 31500.

In addition to SME spectral synthesis, we analyzed the spectrum following a standard spectroscopic method based on the requirements of excitation and ionization equilibria.  Our automated analysis merges the approaches described in \citet{por2008} and \citet{ghe2010}, and will be described in a forthcoming paper of the MARVELS series \citet{fle2011}.  A total of 57 Fe I and 3 Fe II lines were used, after a 2-$\sigma$ clipping was applied to remove those lines with too high or too low iron abundances.  The equivalent widths (EWs) were automatically measured using the task "bplot” in IRAF.  We have derived a $T_{\rm eff} = 5720 \pm 79 $K, $\log{(g)} = 4.20 \pm 0.10$ and metallicity $\rm{[Fe/H]} = -0.37 \pm 0.10$.  We investigated the effects of a contaminating spectrum by modifying the measured EWs by a correction factor equal to the inverse of the expected flux contribution from the primary.  In the case of a solar-type dwarf with a mid-K companion, the expected flux ratio is $\sim$ 10\% in V band, and the EW correction factor is $\sim$ 1.1.  The general effect is to increase the derived effective temperature, surface gravity and metallicity.  However, the increase is small and well within the internal errors of our technique.  We therefore conclude that a flux ratio of 10\% or lower would not significantly affect our measured values based on the EW method. 

We also examine the H$\alpha$ line profile, which serves as a $T_{\rm eff}$ and chromospheric activity indicator, following \citet{lyr2005} and derive a best-fit $T_{\rm eff} = 5461 \pm 118$ K.  This temperature is lower than the value determined via the Fe line EW method, but is affected by contamination from the secondary that fills in the H$\alpha$ wings, resulting in an effective temperature determination that is systematically cooler.  An analysis of the H$\alpha$ core filling suggests that TYC 272 is much more chromospherically active than the Sun, however, there is a degeneracy between flux contamination from a companion and core emission due to activity, both of which can fill in the core of the H$\alpha$ line.  We further note that it is often the case that H$\alpha$-based and color-based $T_{\rm eff}$ are systematically lower than temperatures determined spectroscopically, one possible cause being chromospheric activity \citep{por2008}.  Given the relatively short orbital period of 5.728 days, it is likely that some amount of tidal spin-up has occurred, resulting in stronger magnetic fields and increased starspot activity.

\begin{figure}
\plotone{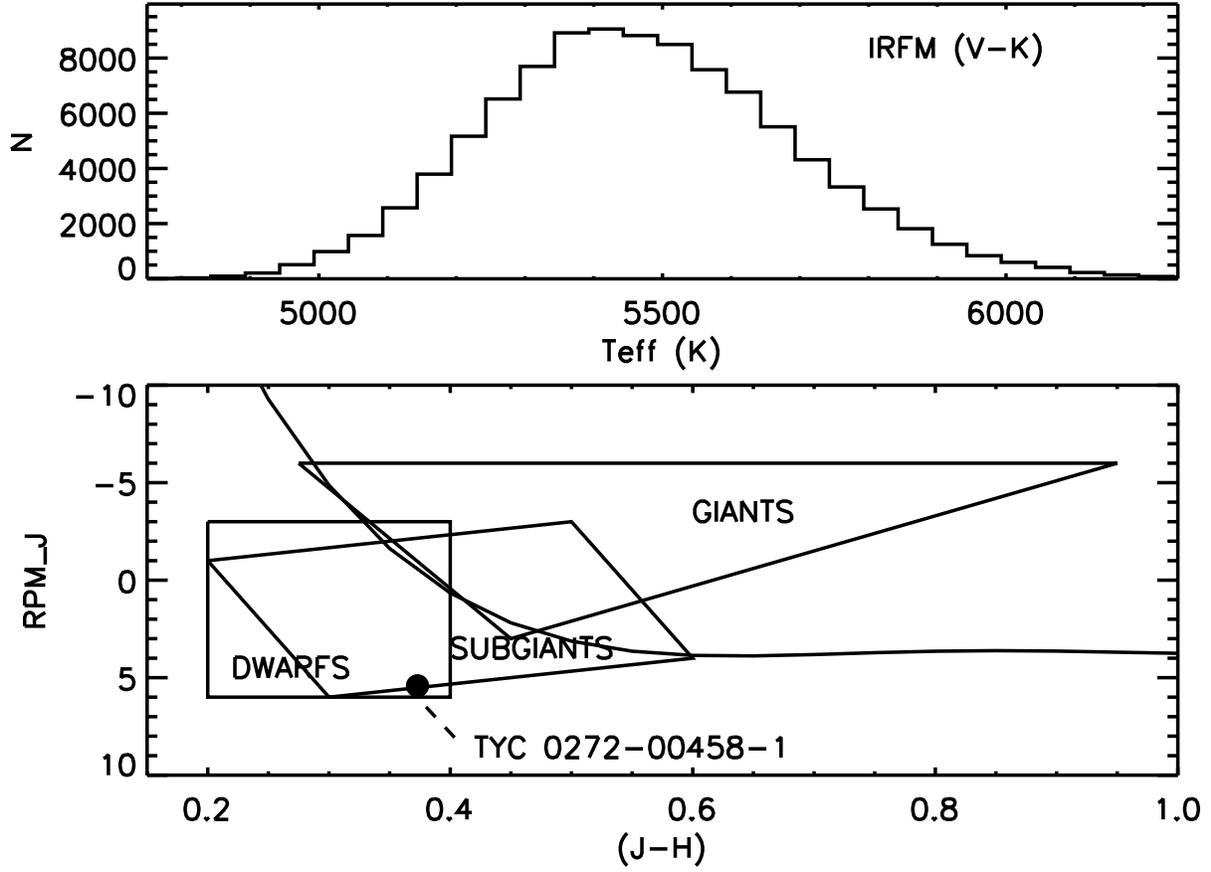}
\caption{Top: Monte Carlo calculation of effective temperature using the infrared flux method.  We estimate a $T_{\rm eff} = 5459 ^{+233} _{-208} $K from this analysis.  Bottom: RPM-J reduced proper motion diagram.  The dividing line between giants and dwarfs is the solid curve.  Approximate regions where dwarfs, subgiants and giants inhabit are labeled.  The location of TYC 272 is consistent with a dwarf or subgiant.\label{irfm_rpmj}}
\end{figure}

We complement the spectroscopic analysis with a variety of additional techniques to estimate stellar parameters based on other observables.  To further estimate the luminosity class, we use a reduced proper motion (RPM) diagram \citep{col2007b} to determine whether the host star is a likely giant or a dwarf/subgiant.  The RPM diagram is excellent at distinguishing giants from dwarfs/subgiants.  We tested the RPM technique on the Kepler Input Catalog (KIC)\footnote{\url{http://archive.stsci.edu/kepler/kic10/search.php}}, which has had its determinations of luminosity class verified using Kepler's high-precision photometry \citep{koc2010}.  We find that only 2\% of KIC giants were classified as dwarfs/subgiants using our RPM analysis.  We use the proper motion values from the Tycho-2 catalog and NIR photometry from 2MASS (Table \ref{tyc0272_starprop}) to place TYC 272 on the RPM-J diagram (Fig.\ \ref{irfm_rpmj}, bottom).  The solid, curved line in the RPM diagram is the dividing line between giants and dwarfs.  The labels show the approximate regions where dwarfs, subgiants and giants are located.  Given how precise the RPM diagram is at identifying giant stars, the fact that TYC 272 lies in the dwarf/subgiant region of the RPM diagram means it is highly unlikely that TYC 272 is a giant star.

We use the infrared flux method (IRFM) \citep{cas2010} to estimate the effective temperature of the primary star via a Monte Carlo approach.  We sample the observed $V$, $J$ and $Ks$ magnitudes assuming Gaussian errors, sample uniformly in [Fe/H] from $-1.0 < \rm{[Fe/H]} < 0.5$ and uniformly in extinction from $0 < A_V < 0.075$, where the upper bound is determined from the Schlegel et al. dust maps \citep{sch1998}.  Fig.\ \ref{irfm_rpmj} shows the results of the Monte Carlo IR flux method (top).  We estimate the effective temperature based on the ($V$-$K\!s$) color to be $T_{\rm eff} = 5459 ~ \left(+233, -208\right) ~ $K, where the uncertainties are 1-$\sigma$ equivalents centered on the median of the asymmetrical $T_{\rm eff}$ distribution.  The difference in temperature between the IRFM and spectroscopic results can be caused by flux contribution from the secondary.  Indeed, the peak of the ($J$-$K\!s$)-based temperature distribution is slightly cooler, which can occur because the flux ratio in the NIR bands is larger than in the optical bands.  This IRFM $T_{\rm eff}$ is therefore a lower bound on the true $T_{\rm eff}$ of the primary due to reddening by the companion.

To further study the effective temperature of the primary, we make use of the spectral energy distribution (SED).  We use the optical band measurements from HAO, near-IR measurements via the 2MASS \citep{skr2006} Point Source Catalog and UV measurements from GALEX \citep{mar2005}, which are compiled in Table \ref{tyc0272_starprop}.  NextGen models from \citet{hau1999} are used to construct the theoretical SED.  Based on the results from the spectroscopic and IRFM analysis, we construct a model with $T_{\rm eff} = 5785$, $\log{(g)} = 4.5$ and $\rm{[Fe/H]} = -0.3$ and compare with the observed fluxes.  These stellar parameters were selected as representative values between the spectral synthesis, EW measurements, H$\alpha$ profile analysis and IRFM temperature calculations.  Distance and extinction are free parameters

\begin{figure}
\plotone{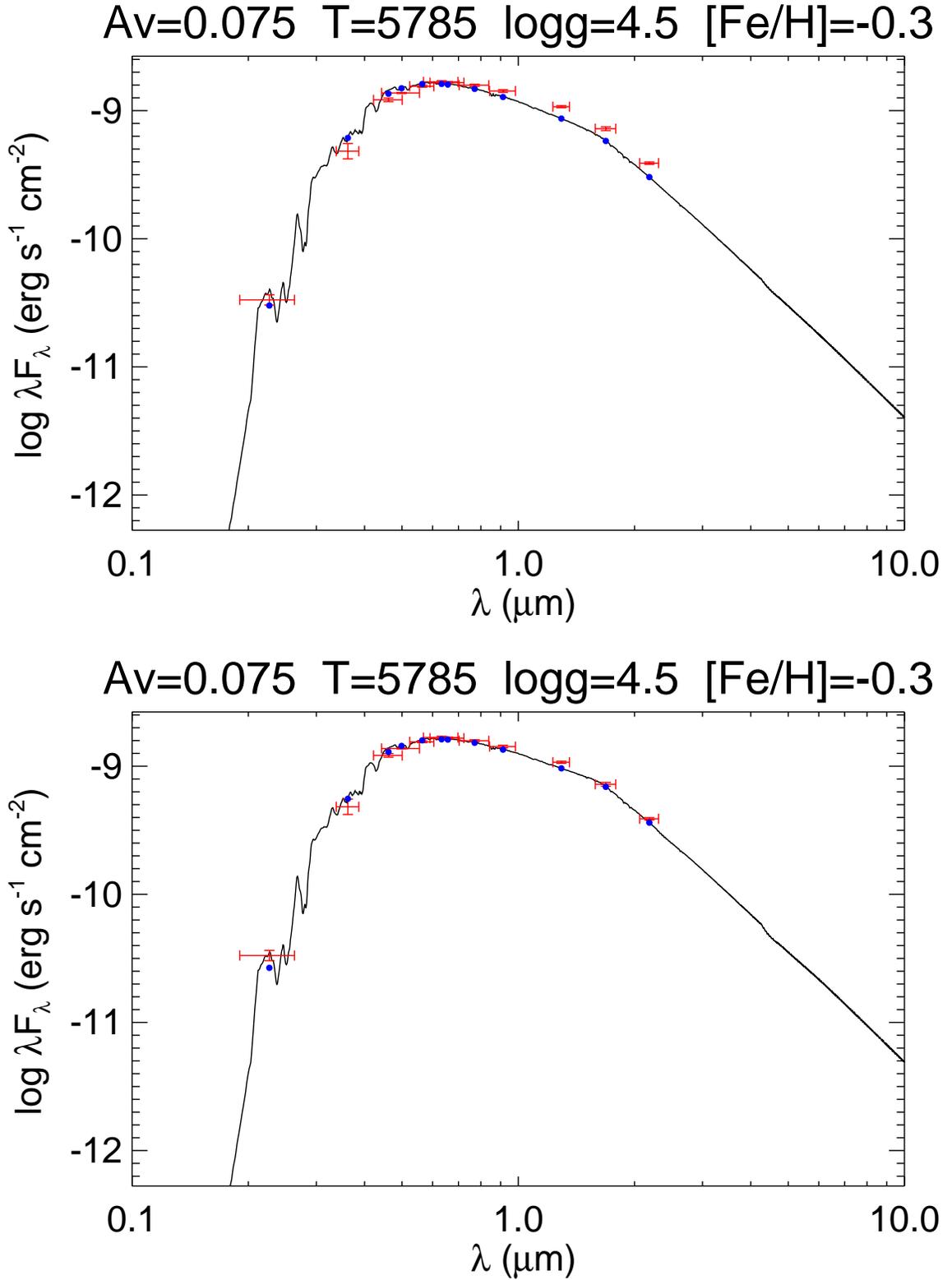}
\caption{SED fits using NextGen models to TYC 272.  Top:  Model that only includes flux from the primary, assuming stellar parameters from the combined spectroscopic analysis.  Bottom:  Model that includes flux contribution from a K5 dwarf secondary, demonstrating the near-IR excess can be explained by a K-type companion.\label{0272sed}}
\end{figure}

Fig. \ref{0272sed} (top) shows the result of the SED fit and the model spectrum.  The blue points are the passband-integrated model fluxes for each filter, while the red crosses are the bandpasses of each filter (horizontal bars) and the measured error in the fluxes (vertical bars).  A clear excess in the near-IR is seen when we do not consider flux from the secondary.  Adopting a primary mass of $1 ~ M_{\odot}$ and a radius of $1 ~ R_{\odot}$, we calculate a secondary mass of $0.64 ~ M_{\odot}$ (K5 spectral type dwarf).  We then add flux contribution from a second star with $T_{\rm eff} = 4400$K and a radius of $0.7 ~ R_{\odot}$, and find that the inclusion of flux from a K-type companion can explain the near-IR excess seen in the SED Fig. \ref{0272sed} (bottom).  This SED analysis does not provide an independent measurement of the stellar parameters, but rather serves as a plausibility experiment that demonstrates flux contribution from a K-type companion can explain the apparent NIR excess if the primary has a $T_{\rm eff} \sim 5785$ K.

\begin{figure}
\plotone{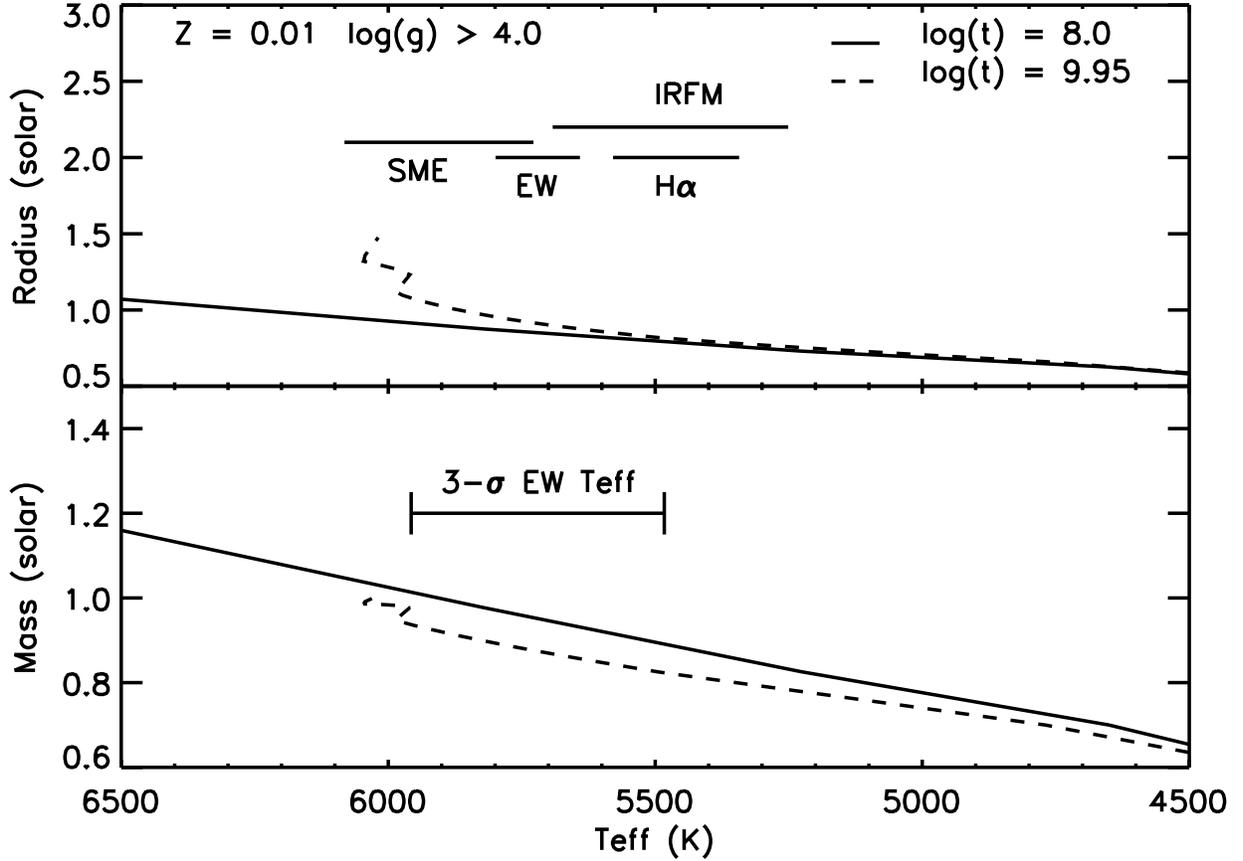}
\caption{Radius (top) and mass (bottom) as a function of effective temperature for different Padova isochrones and metalicity $Z = 0.01$.  The solid line is for an assumed young age (100 Myr) while the dashed line is for an assumed old age ($\sim$9 Gyr).  The horizontal, solid lines are the $T_{\rm{eff}}$ 1-$\sigma$ confidence intervals for the various stellar parameter techniques that were applied.  Only stars with $\log{(g)} > 4.0$ are shown, since there are no indications the primary has a lower $\log{(g)}$.  We use these isochrones to constrain the mass of the primary ($M_1$) in order to derive $M_2$, $R_1$ and $R_2$.\label{0272isochrones}}
\end{figure}

Because the secondary is too faint to have its spectral features detected in the MPP spectra, the mass of the primary cannot be directly measured via RV.  Instead, we make use of Padova isochrones \citep{mar2008} to estimate the mass of the TYC 272 primary.  Fig. \ref{0272isochrones} plots radii and masses as a function of effective temperature for two assumed ages (100 Myr and $\sim$9 Gyr) and a metalicity $Z = $0.01 $\sim \rm{[M/H]} = -0.28$.  This metallicity is chosen based on the SME and EW determinations, which are in good agreement with each other.  The uncertainty in $\rm{[M/H]}$ does not significantly affect the derived range of probable masses for TYC 272, therefore we elect to use a single value for the tracks' metallicities, for clarity.  The derived $T_{\rm eff}$ values from the SME, EW, H$\alpha$ and IRFM are plotted as horizontal bars representing the 1-$\sigma$ confidence intervals.  We elect to use a $T_{\rm eff}$ range equal to the EW 3-$\sigma$ confidence interval $\left(5483 < T_{\rm eff} < 5957~\rm{K}\right)$ as our best estimate for the primary $T_{\rm eff}$ because it is expected to be minimally affected by flux contribution from the secondary.  We note however that all four $T_{\rm eff}$ are consistent to within 2-$\sigma$, and in fact the SME, EW and IRFM temperatures all agree to within 1-$\sigma$.  Based on this $T_{\rm eff}$ range, we find the Padova models constrain $M_1 = 0.92 \pm 0.1 M_\odot$ for $\log{(g)} > 4.0$ and $8.0 < \log{(\rm{age})} < 9.95$.

\subsubsection{Light Curve Parameters}
\label{0272phot}
\begin{figure}
\plotone{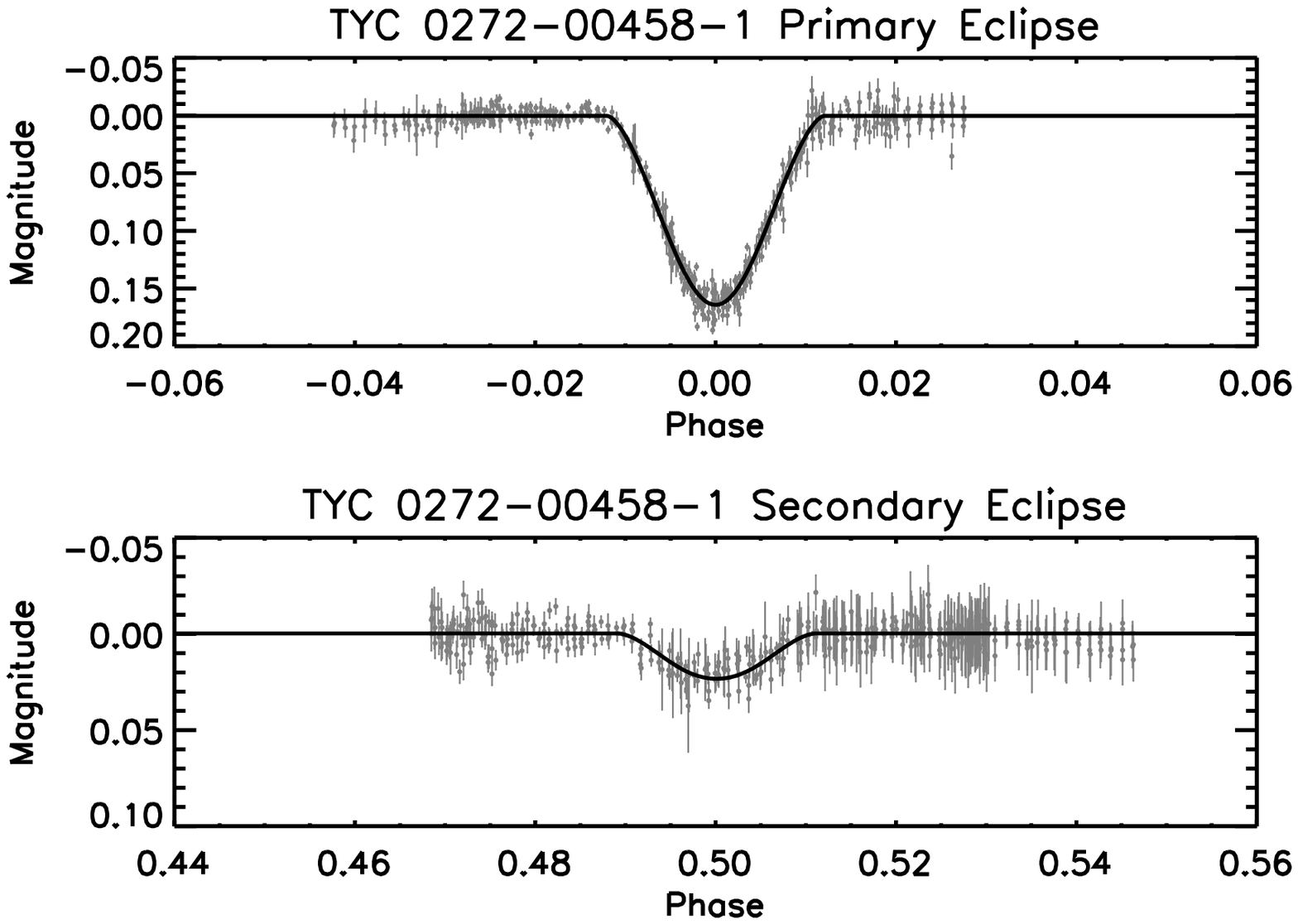}
\caption{Phase-folded SuperWASP lightcurve and best-fit model for the primary (top) and secondary (bottom) eclipses of TYC 272.\label{0272lc}}
\end{figure}

\begin{figure}
\plotone{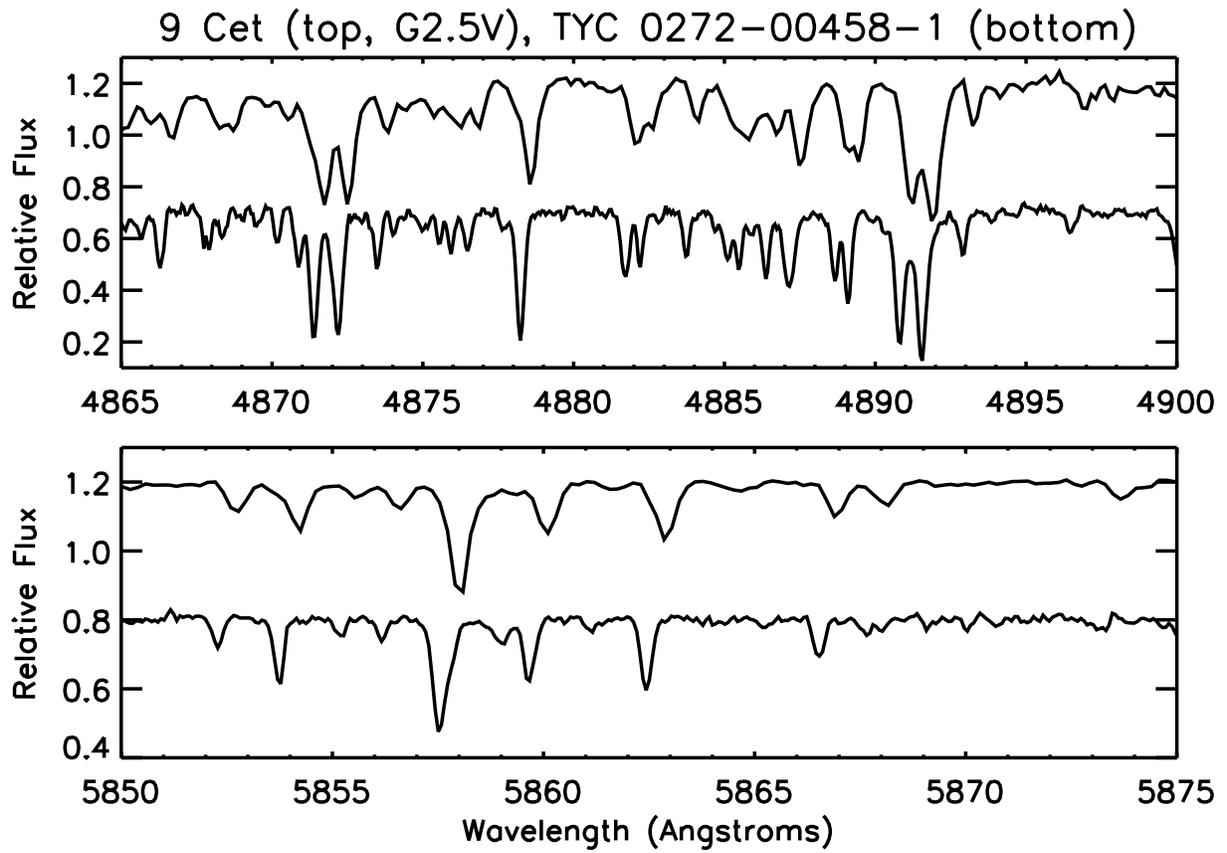}
\caption{Comparison between the R $\sim$ 12,000 FOE spectrum of the G2.5V standard 9 Cet (top of each plot) with the R $\sim$ 30,000 ARCES spectrum of TYC 272 (bottom of each plot).  The two wavelength ranges contain absorption lines present in stars mid-F though mid-M.  No secondary set of spectra can be seen.\label{sb2check}}
\end{figure}

The lightcurve for TYC 272 is shown in Fig.\ \ref{0272lc}, with the best-fit model overplotted as the solid line.  The top panel contains the primary eclipse, while the shallowness of the secondary eclipse can be seen in the bottom panel.  The secondary eclipse of TYC 272 appears to be grazing, leading to model degeneracies between the ratio of the radii and the surface brightness ratio.  This is a well-known problem for grazing EBs, and the solution is to constrain the flux ratio when fitting the lightcurve \citep{and1983}.  The flux ratio is usually constrained spectroscopically \citep{cla2003,sou2004,bla2008}, and we utilized the ARCES spectrum to search for any evidence of a secondary spectrum.  The ARCES spectrum was obtained at an orbital phase corresponding to an expected velocity shift of -27.25 $\rm{km ~ s^{-1}}$, however, the signal-to-noise was not sufficient to detect a secondary spectrum for all flux ratios.  Fig.\ \ref{sb2check} compares the ARCES spectrum with the (lower-resolution, R=12,000) FOE \citep{mon1997,mon1998} spectrum of 9 Cet, a G2 V standard.  Two different wavelength regions are examined that contain absorption features present in stars of mid-F through mid-M spectral types.  No set of secondary spectra are seen, and we estimate an upper limit of $< 20$\% for the flux ratio at these wavelengths.  In the absence of sufficiently high resolution spectra, we estimate the flux ratio using the mass range expected for $M_1$ in \S \ref{0272starprops} and the measured RV semiamplitude of the primary $K_1$.  The flux ratio is constrained to be $8 \pm 4$\%, the estimated flux ratio in $V$ of a G-type primary and K-type companion.  As shown in the SED analysis, such a flux ratio reproduces the observed SED well, particularly in the 2MASS NIR bands.

To model the lightcurves, we used the jktebop\footnote{\url{http://www.astro.keele.ac.uk/~jkt/codes/jktebop.html}} implementation of the EBOP lightcurve model \citep{nel1972,pop1981} to produce a least-squares fit to the data. We modified jktebop to include the normalization for each night of data as additional free parameters. We consulted various tabulations of linear limb darkening coefficients \citep{van1993, cla2000, cla2004} and decided to adopt a limb darkening coefficient of 0.6 in the WASP band for the G-type primary of TYC 272. For the K-type secondary star we used a linear limb darkening coefficient of 0.7. An uncertainty of $\pm$0.1 is assumed for these parameters, and the small effect of this uncertainty is accounted for in the standard errors quoted for other parameters of the lightcurve solution.  The values and uncertainties of $e\cos{\omega}$ and $e\sin{\omega}$ from the RV orbital solution are included as constraints to the least-squares fit to the lightcurve.  Without including such a constraint, the best-fit value of $e\sin{\omega} = 0.2 \pm 0.07$, however, because this is determined via the widths of the eclipses, we do not get a robust estimate from the lightcurve data alone.

Table \ref{0272lcparams} summarizes the geometric parameters derived from fitting the photometry.  To account for correlated errors (``red noise'') in the photometry, uncertainties in the parameters are derived using a ``prayer-bead'' bootstrap Monte Carlo method.  This method uses the residuals from the best fit to create synthetic lightcurves with noise characteristics that are similar to the real data via circular permutation.  Comparison of the parameters that are common to both the RV analysis and the photometric analysis show excellent agreement.  The orbital period agrees to within 1-$\sigma$, while the $e\cos{\omega}$ and $e\sin{\omega}$ values are consistent with a circular orbit.

\subsubsection{Determination of Mass and Radius}
\label{mreqn}
The mass of the secondary, as well as the radii of the two binary components, can be determined in an SB1 if a primary and secondary eclipse are both observed and the mass of the primary star is known (or assumed).  This is accomplished via the mass function:
\begin{equation}
\frac{\left(M_2\right)^3} {\left(M_1 + M_2\right)^2} = \frac{4\pi^2\left(a_1\right)^3} {GP^2}
\label{eqn:massfunc}
\end{equation}

Where $G$ is the gravitational constant, $P$ is the orbital period, and $a_1$ is determined observationally using the definition of the RV semiamplitude $K_1$ from the radial velocity equation:
\begin{equation}
a_{j} = \frac{{K_j ~ P ~ \sqrt[]{1-e^2}}} {{2~\pi~\sin{\left(i\right)}}}, j=\{1,2\}
\label{eqn:rveqn}
\end{equation}
where $e$ is the orbital eccentricity, $i$ is the line-of-sight orbital inclination and the subscript $j = \{1,2\}$ represents the primary and secondary, respectively.  The radii of the two stars can be calculated directly from the geometric lightcurve parameters $R_{1}~a^{-1}$ and $R_{2}~a^{-1}$ found in Table \ref{0272lcparams} and the semimajor axis $a = a_1 + a_2$.

The uncertainties for $M_1$, $M_2$, $R_1$ and $R_2$ are determined as follows.  The uncertainty of $M_1$ is taken from the independent measurement of the primary's mass, if such a determination exists, or from the range of estimates for $M_1$ based on the star's color and/or spectra.  Using implicit partial differentiation, the propagation of error equation, and removing the cross terms since $M_{1}$ and $C$ are independent, the uncertainty in $M_2$ from Eqn. \ref{eqn:massfunc} is given by:
\begin{equation}
\sigma_{M_2} = \left(\frac{4\sigma_{M_1}^2C^2\left(M_1+M_2\right)^2 + \sigma_C^2\left(M_1+M_2\right)^4} {3\left(M_2\right)^2 - 2C\left(M_1 + M_2\right)}\right)^{1/2}
\label{eqn:sigmam2}
\end{equation}
where $C$ is the right-hand-side of Eqn. \ref{eqn:massfunc}:
\begin{equation}
C = \frac{4\pi^2\left(a_1\right)^3} {GP^2}
\label{eqn:rhs}
\end{equation}
and $\sigma_C$ is found by using the propagation of error equation and assuming $P$, $K_1$, $e$ and $i$ are all independent.  Likewise, the uncertainties in the stellar radii are calculated via the propagation of error equation using the measured uncertainties in $R_{1}~a^{-1}$, $R_{2}~a^{-1}$ (Table \ref{0272lcparams}) and the calculated $\sigma_a$ found via propagating the uncertainties in $a_1$ and $a_2$.

\begin{figure}
\plotone{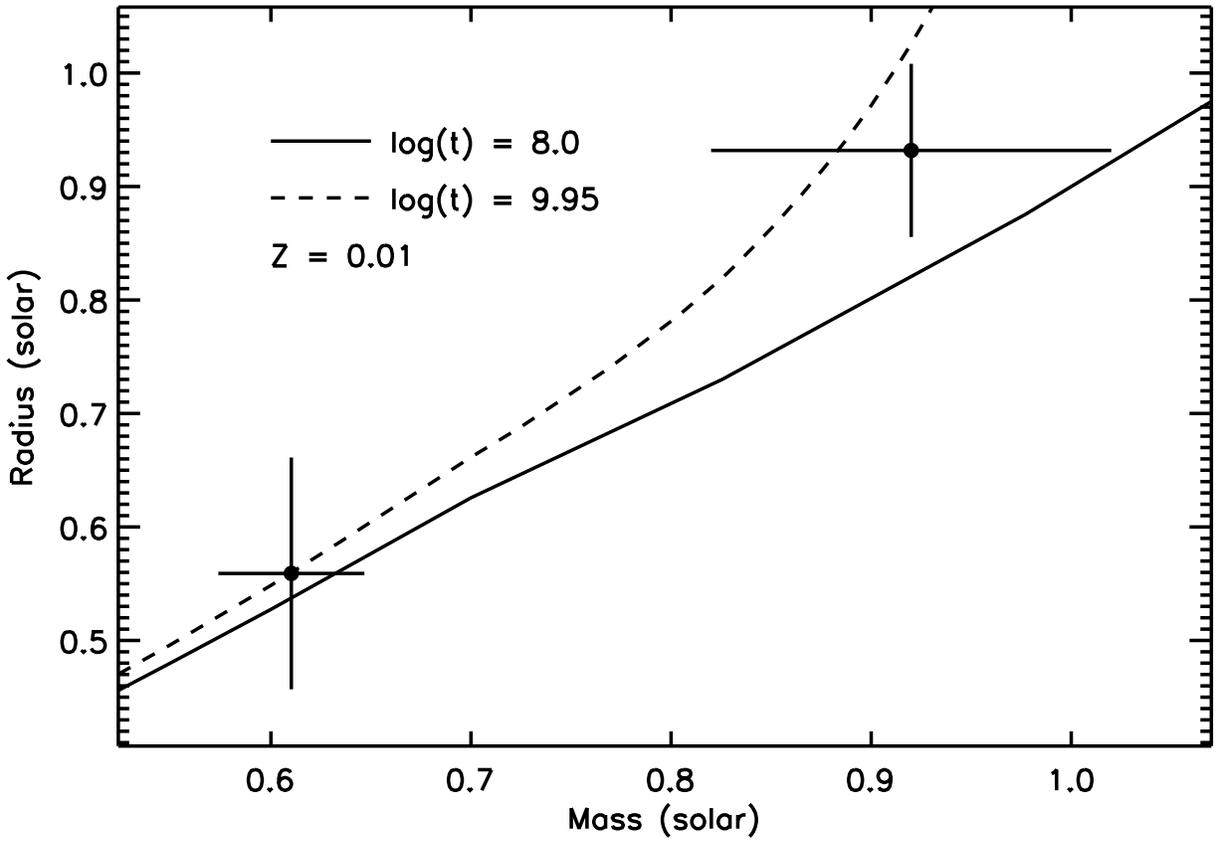}
\caption{Radii and masses for the two components of TYC 272.  Both stars are consistent (1-$\sigma$) with model predictions.  Only a metalicity of $Z = 0.019$ is shown, since it has only a minor effect on the masses and radii of dwarfs and subgiants.\label{0272mandr}}
\end{figure}

We adopt a primary mass of $M_1 = 0.92 \pm 0.1 M_\odot$ based on the stellar characterization presented in \S \ref{0272starprops} and calculate $M_2 = 0.610 \pm 0.036 ~ M_{\odot}$,  $R_1 = 0.932 \pm 0.076 ~ R_{\odot}$ and $R_2 = 0.559 \pm 0.102 ~ R_{\odot}$ (Fig. \ref{0272mandr}).  Both the primary and secondary are consistent (1-$\sigma$) with the model predictions for mass and radii.  Metalicity is a minor effect on the mass and radius of the isochrones, so only the tracks corresponding to $Z = 0.01$ are shown.

Although the relative errors for the masses and radii in the specific case of TYC 272 are larger than typically used to study the mass-radius relationship ($\sim6$\% for $M_2$, 8\% and 18\% for $R_1$ and $R_2$, respectively), relative errors of a few percent are possible if better constraints on the primary mass could be determined and more precise photometry of the grazing secondary eclipse could be obtained.  Specifically, large-aperture telescopes and multi-dimensional correlation software \citep{zuc1994,zuc2003} can be used to extract RV solutions for faint companions, such that the masses of both components can be measured without relying on models or empirical relations to constrain $M_1$.  In addition, high precision photometric follow-up of the shallow secondary eclipse will improve the precision in the radius.  TYC 272 is not an ideal benchmark system because of the shallow, grazing secondary eclipse, however, the analysis presented here demonstrates the required steps to derive precise masses and radii for EBs using a combination of transit and RV exoplanet survey data.  A summary of these analysis steps for EBs with low flux ratios, such as TYC 272, are described in \S \ref{framework}.

\subsection{TYC 1422}
\subsubsection{Orbital Parameters}
TYC 1422 had a total of 16 epochs observed during the MPP, with a $Q$ = 3.07 for its field and an average, scaled RV uncertainty of 0.304 $\rm km ~ s^{-1}$.  When fitting the MPP RV measurements for TYC 1422, the residuals were found to be exceedingly large (RMS $\sim 6 ~ \rm{km ~ s^{-1}}$).  Such a large scatter is often caused by contamination of an additional spectrum from an unresolved source, a likely scenario for short-period SBs with similar fluxes.

The MPP DFDI pipeline was not designed to treat multiple sources within a single spectrum, and can produce large systematic errors in those cases.  It is important to note that the inability to obtain precise RV measurements for SB2s applies to the MPP instrument only.  Other survey instruments, including the MARVELS survey instrument itself, will be able to treat the case of double-lined binaries.  The MARVELS survey pipeline stores information about the fringe fitting that can be used to remove the fringes from the stellar DFDI spectrum and re-create a "traditional", de-fringed stellar spectrum.  This spectrum can then be used to perform cross-correlation analysis and identify and measure multiple CCF components, thereby measuring RVs of double-lined binary stars.

Because the MPP instrument passed starlight through an Iodine cell, a similar reconstruction of the stellar spectrum is not possible due to the Iodine lines that are present.  Attempting to model the Iodine lines and remove them from the spectrum prior to de-fringing is also difficult, because the Iodine lines are numerous and very narrow.  In fact, most of the Iodine lines are unresolved at the resolution of the MPP instrument.  For this reason, independent RV follow-up observations (SMARTS data) were obtained to measure the RVs of the multiple components.  In general, such additional follow-up is not required and RV measurements for both components of double-lined binaries can be obtained directly from the survey data itself.

Despite the large scatter in the residuals, the derived orbital period from the MPP data is in good agreement with the period derived from the SuperWASP photometry ($\sim$ 6.2 days).  To derive an improved orbital solution, we use the higher resolution SMARTS spectra to resolve the components and measure the Doppler reflex velocities of the individual components.  We do not use the MPP data in our final analysis because it is so heavily affected by systematic errors.  However, for completeness, we present the MPP RV data in Table \ref{1422rvs}.

\begin{figure}
\plotone{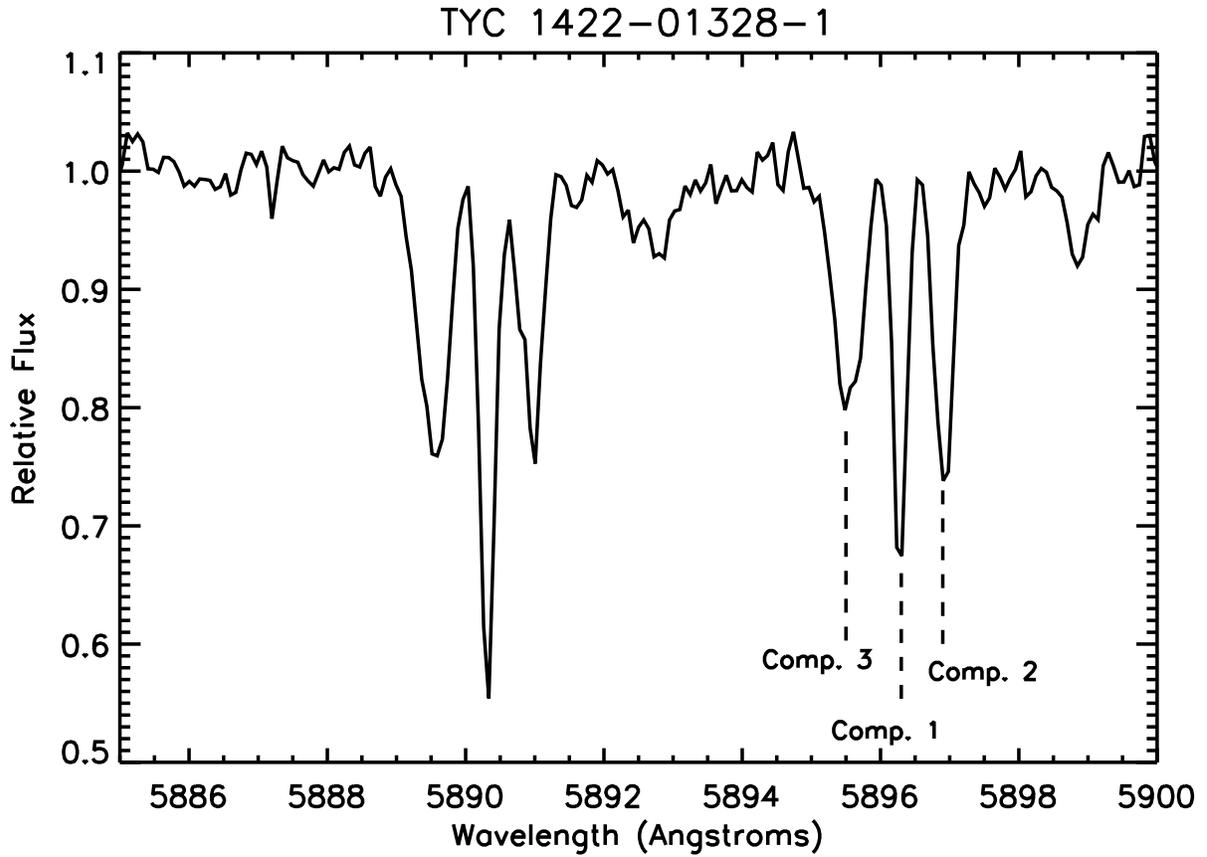}
\caption{ARCES spectrum of TYC 1422 centered on the Na D doublets.  The three sets of doublets can be clearly seen, and components are marked with the same names assigned to them while measuring their RVs with the SMARTS data.  The components were identified in the SMARTS CCF's by examining the width and shape of their CCF peaks.\label{sb3detect}}
\end{figure}

\begin{figure}
\plotone{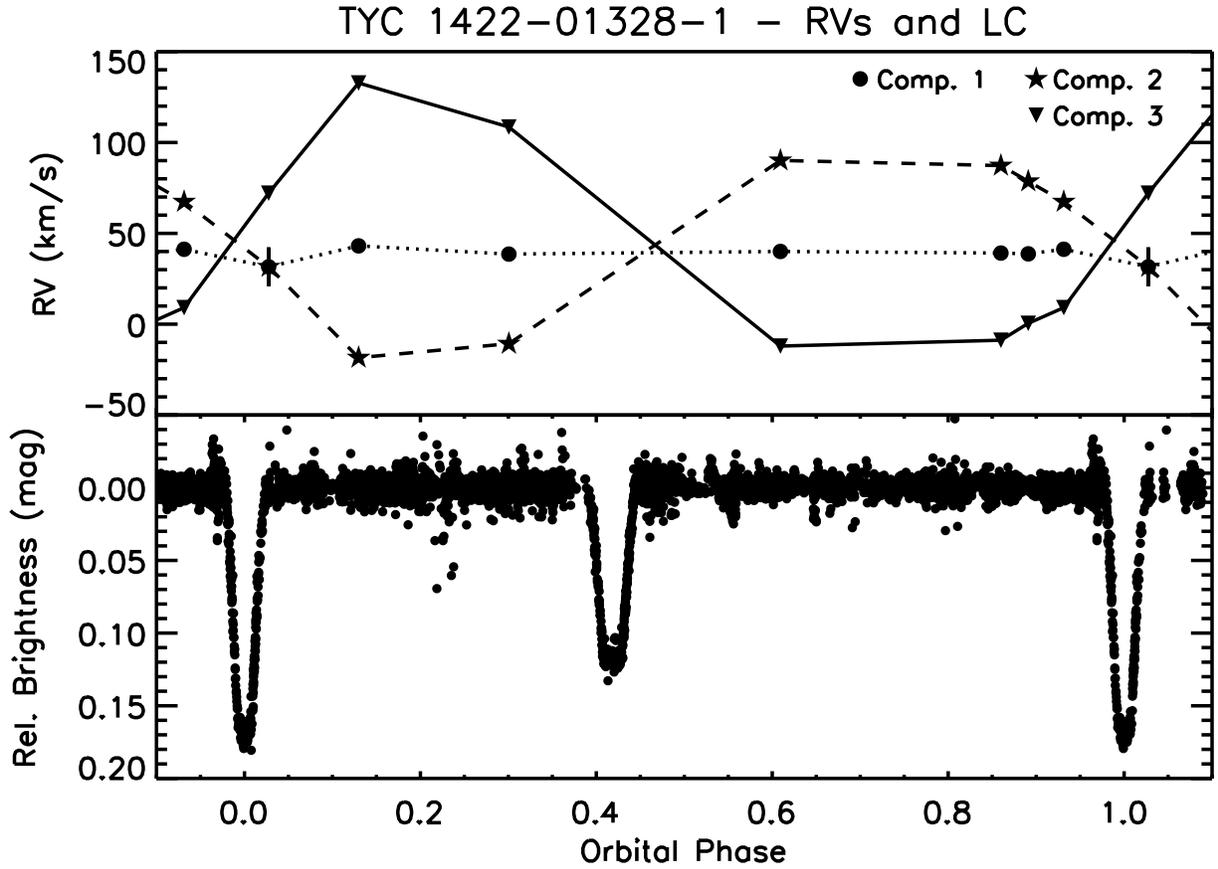}
\caption{Photometry and RVs for the three components of TYC 1422 phase-folded to the best-fit period and epoch of central transit in Table \ref{1422lcparams}.  The connecting lines between the measured RVs for each component serve as a visual aid.  Components 2 and 3 are identified as the EB pair.  Their measured RVs are antiphased as one would expect for an orbiting pair.  They also phase appropriately such that the RV zero-crossing occurs near primary and secondary eclipses.  Note the connecting lines do not cross at the precise primary and secondary eclipse phases because they do not represent the actual Keplerian orbit.\label{compphase}}
\end{figure}

\begin{figure}
\plotone{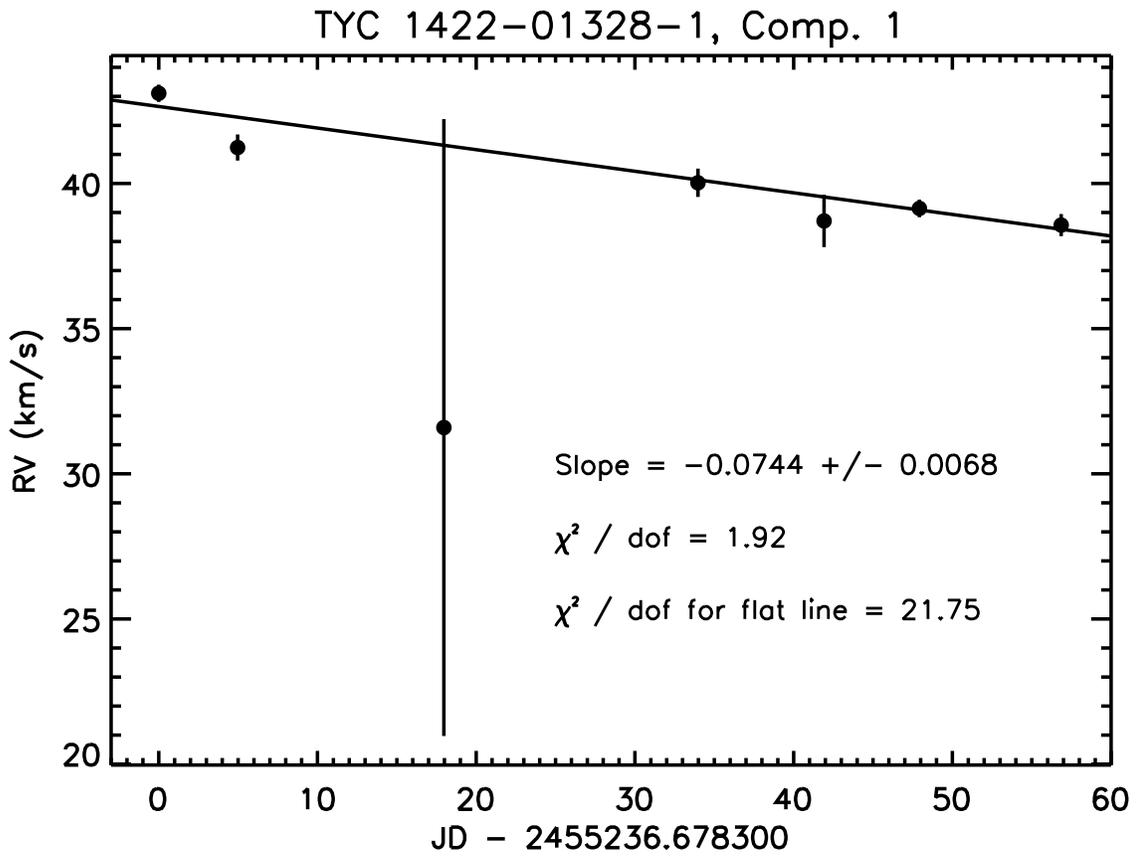}
\caption{RVs for Component 1 in TYC 1422.  The large error bar for the third epoch is caused by blending.  The data are well-fit with a gradual slope of $\sim -75 ~ \rm{m ~ s^{-1} ~ day^{-1}}$, which could be residual instrument drift given that a similar slope is found when fitting Components 2 and 3.\label{comp1rv}}
\end{figure}

The TYC 1422 system consists of three sources, as demonstrated in the triple set of Na D doublets from the ARCES spectrum in Fig.\ \ref{sb3detect}.   We do not attempt an SED analysis due to the complexity of the system, however, the basic stellar properties, including the HAO absolute photometry, are compiled in Table \ref{tyc1422_starprop}.  The RVs for each component are phase-folded to the orbital period and ephemeris as determined by the SuperWASP photometry in Fig.\ \ref{compphase}.  The RV measurements for each component are connected by different line types for visualization.  Components 2 and 3 are identified as the eclipsing binary pair.  Their RVs are in antiphase, as expected for an orbiting pair, and have relative velocities that cross the zero-line near the phases of primary and secondary eclipse.  Component 1 is found to have a gradual linear trend of $\sim ~ -75 ~ \rm{m ~ s^{-1} ~ day^{-1}}$.  Fig.\ \ref{comp1rv} shows the RVs from Component 1 and the best-fit linear relation.  The large uncertainty for the third epoch is caused by blending.  Table \ref{1422smartsrvs} contains the times and measured RVs for each of the three components.

\begin{figure}
\plotone{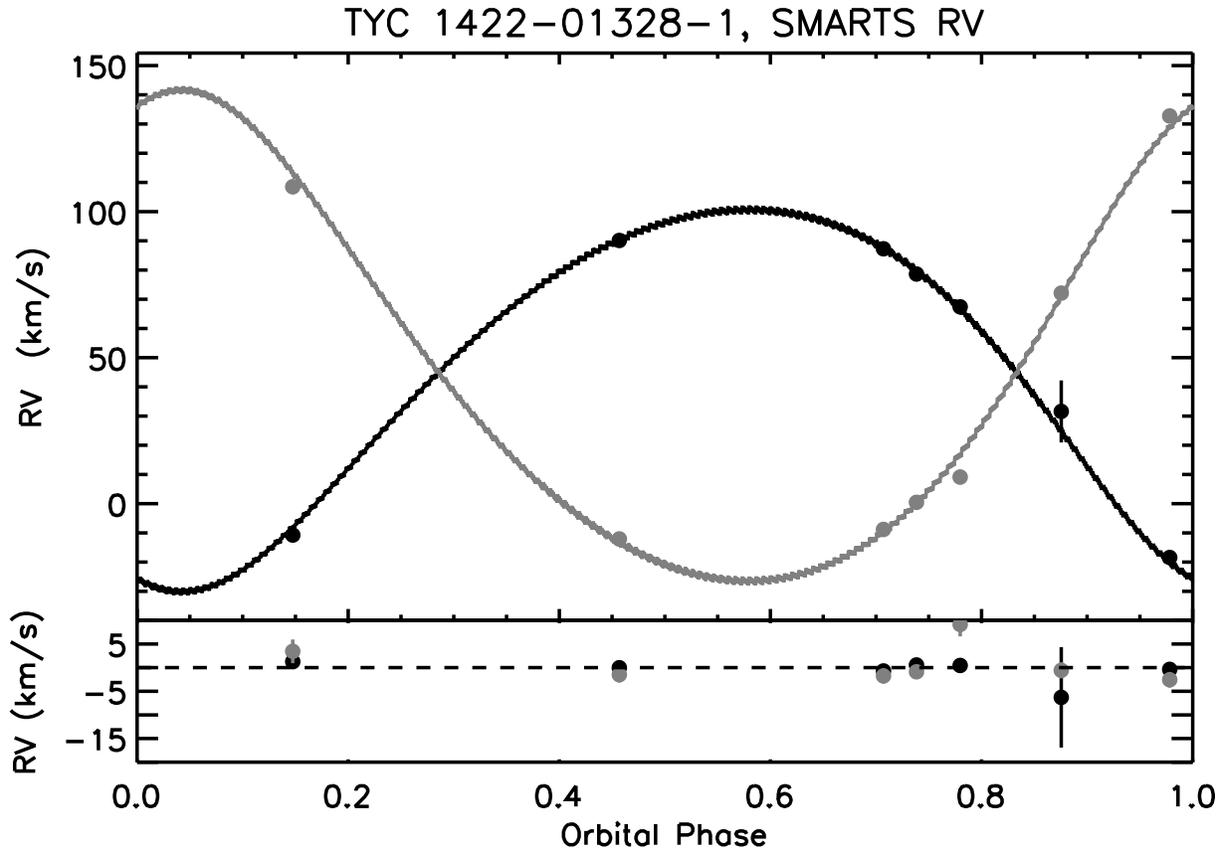}
\caption{Best-fit orbital solution for Components 2 and 3 of TYC 1422 from the SMARTS observations.  The black data points and curve are the RVs for Component 2 and the best-fit orbital solution, while the grey points are for the less massive Component 3.  The jagged nature of the best-fit curve is a result of phase-folding the small linear-trend included in the fitting.  The bottom panel shows the residuals to the fit.\label{comp23rv}}
\end{figure}

The SMARTS RVs were fit using MPRVFIT2, an IDL procedure written by N. De Lee which combines a periodogram searching algorithm and a non-linear least squares fitter to find the optimal Keplerian orbit for a given set of radial velocity points.  The period search algorithm consists of a modified Lomb-Scargle periodogram based on \citet{cum2004} with an analytical False Alarm Probability (FAP) derived from \citet{bal2008}.  The software allows for simultaneous fitting of RVs from both components of a binary system.  The SMARTS spectra consist of seven epochs, resulting in only one degree of freedom for a full Keplerian fit.  However, because Components 2 and 3 are a physically bound pair, we can exploit the fact that several parameters are identical for the two components.  Specifically, the period ($P$), eccentricity ($e$), epoch of periastron ($T_{p}$) and systemic RV ($\gamma$) are identical, while the argument of periastron ($\omega$) differs by $\pi$ radians.  This results in seven degrees of freedom between the 14 combined data points, the seven parameters being $\left\{P,e,\omega,T_{p},K_{1},K_{2},\gamma\right\}$, where $K_{1}$ and $K_{2}$ are the RV semi-amplitudes of the two components.  Due to the linear trend observed in Component 1, we also include a linear term that reduces the degrees of freedom to six.  Fig.\ \ref{comp23rv} shows the best-fit orbital solution for Components 2 and 3, and the orbital parameters are presented in Table \ref{tyc1422_dynprop}.  We find a best-fit period of 6.2005 days and a slightly eccentric orbit with $e = 0.16$ and a mass ratio of 0.78.  The orbital period is in agreement with the photometric period to within $1-\sigma$.  The slope of the linear term is $-48 \pm 18 \rm{m ~ s^{-1} ~ day^{-1}}$, which is of comparable magnitude and direction as the slope measured on Component 1, and therefore may be residual instrument drift.

\subsubsection{Light Curve Parameters}
We performed the same analysis using the jktebop software as done for TYC 272, however, third light (flux contributed from Component 1, $L_3$ in Table \ref{1422lcparams}) is included as a free parameter in the least-squares fit to the lightcurve of TYC 1422.  In principle one can compare a spectroscopic flux ratio based on the EWs of spectral lines with the best-fit flux ratio derived from the lightcurve solution.  However, in the case of TYC 1422 and the ARCES spectrum, most of the lines are blended at the resolution of the spectrograph.  The Na D lines shown in Fig.\ \ref{sb3detect} are quite visible, however, the wings of the features are still blended, and it is challenging to find an appropriate pseudo-continuum from which to measure the EWs.  Furthermore, the EWs of the Na D doublets are known to depend on $T_{\rm eff}$ \citep{tri1997, dia2007}, and the $T_{\rm eff}$ of the third component is not constrained in our analysis.

\begin{figure}
\plotone{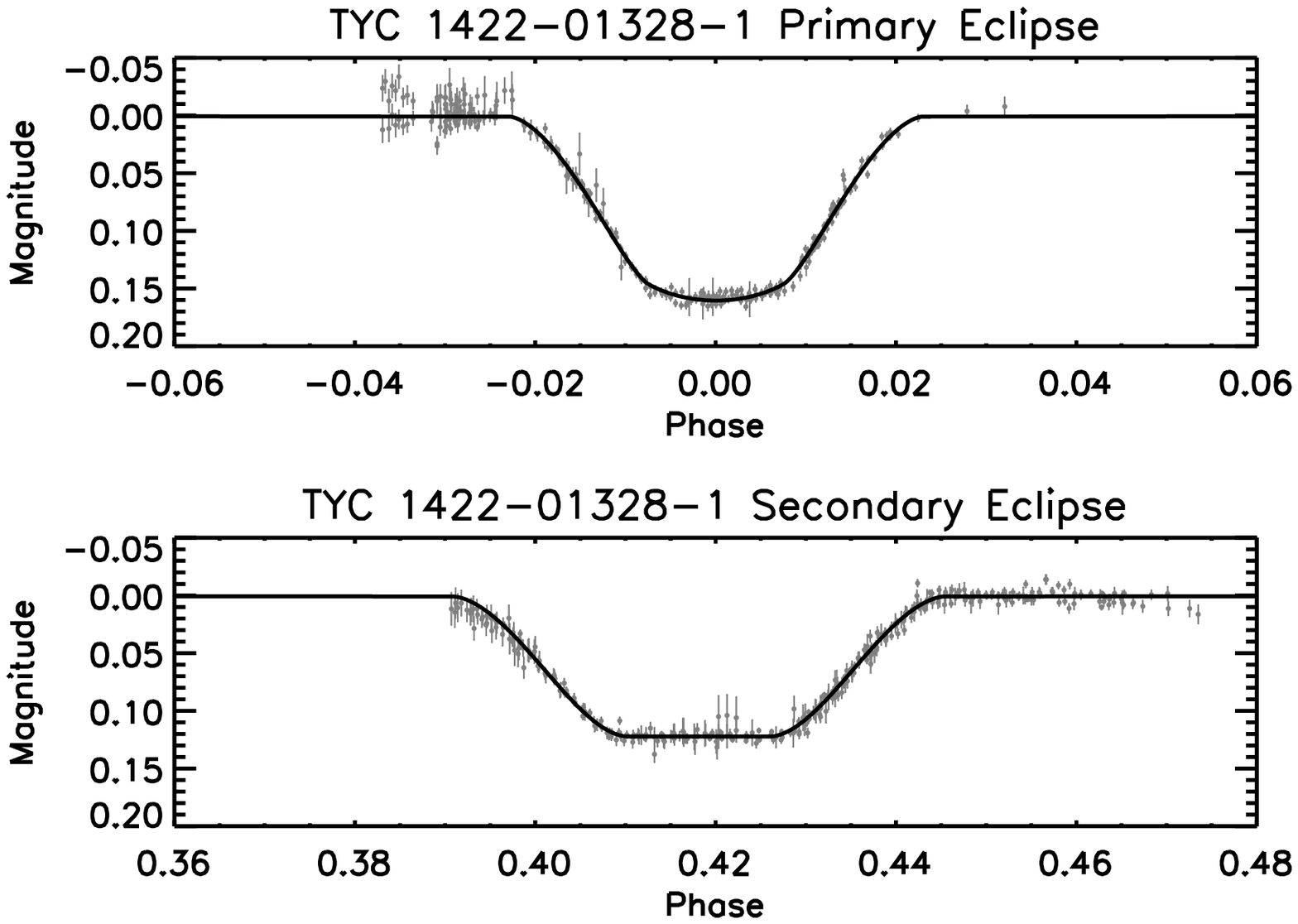}
\caption{Phase-folded SuperWASP lightcurve and best-fit model for the primary (top) and secondary (bottom) eclipses of TYC 1422.\label{1422lc}}
\end{figure}

We use limb darkening coefficients of 0.55 for the primary and secondary.  The third light is accounted for within the WASP band, which ranges from 400-700 nm. and is similar to  $g^\prime + r^\prime$.  The lightcurve with the best-fit model overplotted is shown in Fig.\ \ref{1422lc}.  Table \ref{1422lcparams} summarizes the geometric parameters derived from photometry fitting.  Correlated errors are accounted for using the ``prayer-bead'' bootstrap Monte Carlo approach, as described in \S \ref{0272phot}.  Timing of the primary and secondary eclipse events confirm the system is in an eccentric orbit.

\subsubsection{Determination of Mass and Radius}
Since both components of the EB pair have Doppler measurements, we can measure the masses and radii of Components 2 and 3 directly from the RV semi-amplitudes and the lightcurve.  We find the semimajor axes $a_1$ and $a_2$ using the RV semiamplitudes from Eqn. \ref{eqn:rveqn}.  The masses can then be determined using the center of mass equation and the complete version of Kepler's Third Law:
\begin{equation}
\left(M_1 + M_2\right) = \frac{4~\pi^2~\left(a_1+a_2\right)^3} {G~P^2}
\label{eqn:kep3}
\end{equation}
where $M_1$ and $M_2$ are the masses of the primary and secondary, respectively.

\begin{figure}
\plotone{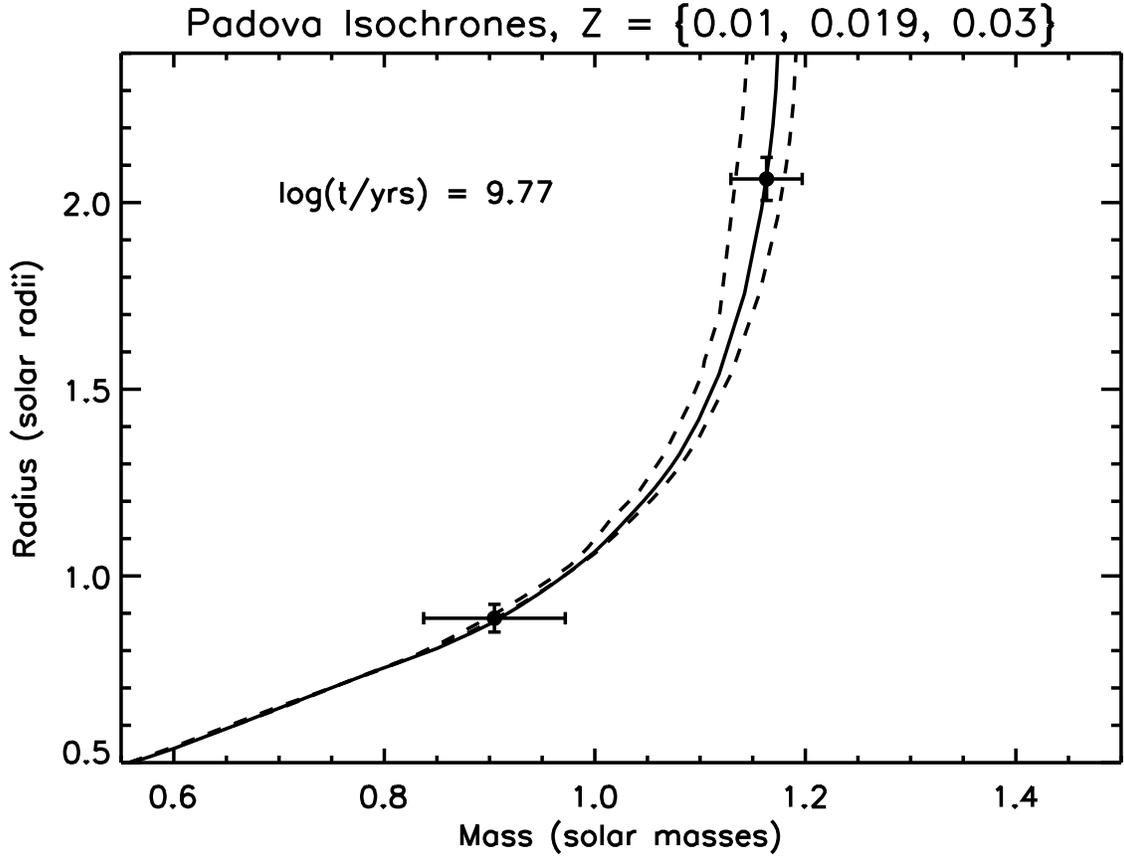}
\caption{Mass-Radius diagram for TYC 1422, Components 2 and 3.  Padova isochrones are overplotted using a $\log{(t)} = 9.77$.  This age was selected as the best-fit by visual inspection of tracks with differing ages.  We find excellent agreement using a solar-metalicity of $Z = 0.019$.  For comparison, we also show tracks with $Z = 0.01$ (left track, dotted line) and $Z = 0.03$ (right track, dotted line).\label{1422mandr}}
\end{figure}

The radii of the two stars can be calculated directly from the geometric parameters measured from the lightcurves in Table \ref{1422lcparams} and the value of the semimajor axis $a = a_1 + a_2$.  The primary star is evolved off the main sequence and has a mass of $M_1 = 1.163 \pm 0.034 ~ M_{\odot}$ and a radius of $R_1 = 2.063 \pm 0.058 ~ R_{\odot}$.  The secondary has a mass of $M_2 = 0.905 \pm 0.067 ~ M_{\odot}$ and a radius of $R_2 = 0.887 \pm 0.037 ~ R_{\odot}$ and is consistent with a main-sequence G-type dwarf.  The relative uncertainties for $R_1$ and $R_2$ are 2.8\% and 4.2\%, very close to the desired precision goal of $<3$\%.  Additional, high precision photometry of the eclipses will be helpful to further improve the precisions.  The relative uncertainties for $M_1$ and $M_2$ are 2.9\% and 7.4\%.  The larger uncertainty for $M_2$ is in part due to the larger uncertainties in the RV measurements for that component.  The SMARTS spectra had relatively low signal-to-noise ($\sim 20$ per resolution element), and obtaining higher signal-to-noise spectra will reduce the uncertainty in $M_2$ to better than 3\%.  Fig. \ref{1422mandr} compares the masses and radii with Padova isochrones at an age of $\log{(t)} = 9.77$ and metalicities of $Z = \{0.01, 0.019, 0.03\}$.  TYC 1422 is consistent with solar metalicity (solid line), whereas the dashed lines are additional metalicity tracks for comparison purposes.

\section{Framework For EBs From Surveys}
\label{framework}
When the RV semiamplitudes for both components can be measured spectroscopically (SB2s), the masses and radii for both stars can be determined directly from observable quantities in the survey data.  TYC 1422 is such an example, although it is not a good benchmark system due to the presence of a third spectrum.  Additional spectra can blend with the other components and affect their derived RVs, and the additional flux contribution complicates the photometry modeling.  Despite these challenges, we were able to obtain 2.9\% relative precision on $M_1$, 2.8\% relative precision on $R_1$ and 4.2\% relative precision on $R_2$.  The MPP data were unable to be de-fringed due to the presence of the Iodine lines from the gas cell, and therefore follow-up observations with SMARTS data were required.  Obtaining better signal-to-noise spectra would reduce the uncertainties in the masses further.  We note that the MARVELS survey instrument does not pass starlight through an Iodine cell, and spectra of SBs with similar flux ratios can be de-fringed and analyzed using cross-correlation techniques like the ones used on the SMARTS spectra of TYC 1422 without additional follow-up.

\begin{figure}
\includegraphics[angle=270,scale=0.6]{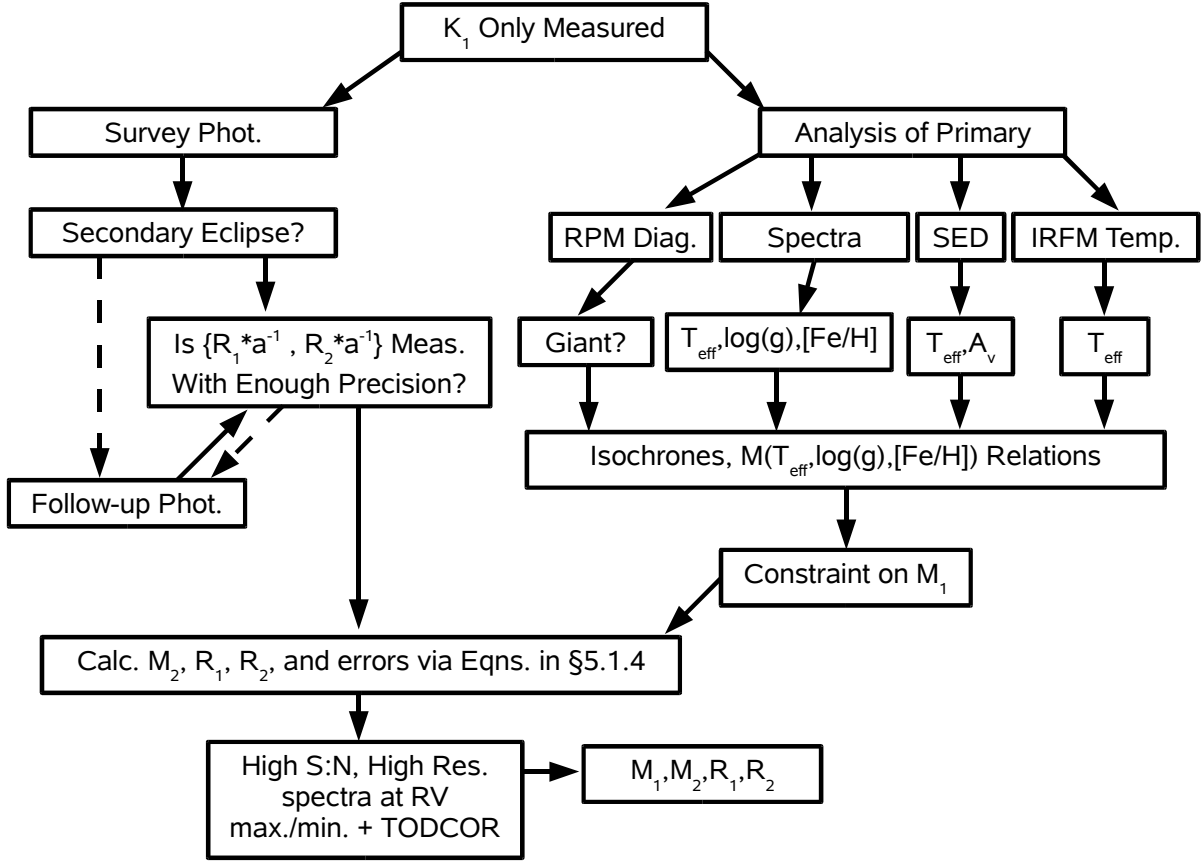}
\caption{Suggested analysis flow diagram that demonstrates the steps one would take to obtain masses and radii for an EB with a small flux ratio, starting with exoplanet survey data and catalog information.  A dashed line represents a ``no'' decision.  The analysis depicted here only works for EBs that undergo both primary and secondary eclipses. \label{workflowdiag}}
\end{figure}

EBs with smaller flux ratios require model-dependent constraints on the primary mass to derive the mass of the secondary and both radii, until observations with larger telescopes can be obtained.  In this paper, we provide a framework for identifying EBs of interest from SB1s using survey data and catalog information.  We summarize the steps taken in a suggested ``analysis flow diagram'' presented in Fig. \ref{workflowdiag}.  This framework applies only to those EBs that have both a measurable primary and secondary eclipse.  Once the mass of the primary and an uncertainty are adopted via comparison with isochrones and/or using empirical relations, the mass of the secondary, both stellar radii, and their uncertainties can be calculated using the equations presented in \S\ref{mreqn}.

We apply this technique to TYC 272, however, the grazing secondary eclipse makes it a poor benchmark EB as well.  The 5.9\% relative error on $M_2$ is dominated by the large uncertainty for the estimate of $M_1$.  Observations with high resolution spectrographs at high signal-to-noise can be used to derive $K_1$ and $K_2$ using TODCOR \citep[see][for an example in the near-IR]{ben2008}.  Such a technique can exploit the fact that the other orbital parameters are well-determined from the survey data, and only a measurement of $K_2$ is required to directly measure $M_1$ and thereby complete the analysis.  In addition, obtaining a high signal-to-noise, high resolution spectrum at a phase where the spectral lines are well-separated can yield spectroscopically-determined flux ratios, which are then used as additional constraints in the lightcurve modeling of eclipses that are grazing.

In favorable cases where there are deep, total eclipses and no third light contamination, transit survey photometry itself can yield radii to better than 2\% relative error \citep{sou2011}.  However, in both SB2 and SB1 situations, the survey photometry may sometimes be insufficient to obtain such high precisions.  Because the epoch of central transit is well-determined from the survey data, follow-up observations can be planned when a primary or secondary eclipse is known to occur.  In addition, obtaining transits in multiple filters can aid in the EB analysis.  We note that this can be a problem for any study of EBs using transiting exoplanet survey data, and is not unique to the methodology presented here.

\section{Conclusion}
\label{conclusion}
We have presented a framework for eclipsing binary science via cross-referencing of photometric and Doppler exoplanet survey data.  We have found two new EBs around F/G primaries; TYC 272, an SB1 consisting of a K dwarf secondary, and TYC 1422, a triple-lined SB that includes an evolved primary star and a G dwarf secondary.  Although neither of these two specific binaries are ideal benchmark systems for studying the mass-radius relationship, we have shown that the combination of survey data can be a rich source of EBs, for which many physical parameters can be directly measured or constrained from the survey data itself.  In particular, the analysis of TYC 272 provides a framework for estimating the masses and radii of SB1s from survey data.  Obtaining additional, high resolution, high signal-to-noise spectra can then directly yield the semiamplitudes, and thus masses, of both components.  Such an approach can be used in particular to study the mass-radius relationship of F/G + K/M EBs.

\acknowledgements
We thank Roger Cohen for useful discussions on theoretical isochrones.  L.G. thanks Dr. Simone Daflon and Dr. Herman Hensberge for the helpful discussions. L.G. acknowledges financial support provided by the PAPDRJ - CAPES/FAPERJ Fellowship.  L. Dutra-Ferreira acknowledges financial support provided by CAPES fellowship.  G.F.P.M. acknowledges financial support from CNPq grants no. 476909/2006-6 and 474972/2009-7, plus a FAPERJ grant no. APQ1/26/170.687/2004.  Funding for the multi-object Doppler instrument was provided by the W.M. Keck Foundation.  The pilot survey was funded by NSF with grant AST-0705139, NASA with grant NNX07AP14G and the University of Florida.  The SuperWASP Consortium consists of astronomers primarily from the Queen's University Belfast, St. Andrews, Keele, Leicester, The Open University, Isaac Newton Group La Palma and Instituto de Astrof\'{i}sica de Canarias. The SuperWASP Cameras were constructed and operated with funds made available from Consortium Universities and the UK’s Science and Technology Facilities Council.  This research has made use of the SIMBAD database, operated at CDS, Strasbourg, France.  Based on observations with the SDSS 2.5-meter telescope.  Funding for the SDSS and SDSS-II has been provided by the Alfred P. Sloan Foundation, the Participating Institutions, the National Science Foundation, the U.S. Department of Energy, the National Aeronautics and Space Administration, the Japanese Monbukagakusho, the Max Planck Society, and the Higher Education Funding Council for England. The SDSS Web Site is http://www.sdss.org/.  The SDSS is managed by the Astrophysical Research Consortium for the Participating Institutions. The Participating Institutions are the American Museum of Natural History, Astrophysical Institute Potsdam, University of Basel, University of Cambridge, Case Western Reserve University, University of Chicago, Drexel University, Fermilab, the Institute for Advanced Study, the Japan Participation Group, Johns Hopkins University, the Joint Institute for Nuclear Astrophysics, the Kavli Institute for Particle Astrophysics and Cosmology, the Korean Scientist Group, the Chinese Academy of Sciences (LAMOST), Los Alamos National Laboratory, the Max-Planck-Institute for Astronomy (MPIA), the Max-Planck-Institute for Astrophysics (MPA), New Mexico State University, Ohio State University, University of Pittsburgh, University of Portsmouth, Princeton University, the United States Naval Observatory, and the University of Washington.  Based on observations obtained with the Apache Point Observatory 3.5-meter telescope, which is owned and operated by the Astrophysical Research Consortium.  This publication makes use of data products from the Two Micron All Sky Survey, which is a joint project of the University of Massachusetts and the Infrared Processing and Analysis Center/California Institute of Technology, funded by the National Aeronautics and Space Administration and the National Science Foundation.  The Center for Exoplanets and Habitable Worlds is supported by the Pennsylvania State University, the Eberly College of Science, and the Pennsylvania Space Grant Consortium.  This work was conducted in part using the resources of the Advanced Computing Center for Research and Education at Vanderbilt University, Nashville, TN.

%tables here
\begin{deluxetable}{rrrr}
\tabletypesize{\scriptsize}
\tablecaption{MPP RV Observations - TYC 272\label{0272rvs}}
\tablewidth{0pt}
\tablehead{
\colhead{HJD$_{\rm{UTC}}$} & \colhead{RV} & \colhead{$\sigma_{RV}$ (formal)} & \colhead{$\sigma_{RV}$ (scaled)}\\
\colhead{~} & \colhead{(km s$^{-1}$)} & \colhead{(km s$^{-1}$)} & \colhead{(km s$^{-1}$)}
}
\startdata
2454101.98701 &  22.075 & 0.086 & 0.297 \\
2454106.01738 & -13.138 & 0.082 & 0.282 \\
2454158.78668 &  -3.698 & 0.100 & 0.343 \\
2454186.75584 & -19.438 & 0.081 & 0.278 \\
2454188.74844 &  70.056 & 0.088 & 0.302 \\
2454189.76185 &  87.249 & 0.090 & 0.310 \\
2454194.79179 &  80.652 & 0.122 & 0.420 \\
2454195.81803 &  79.963 & 0.119 & 0.409 \\
2454215.66891 & -16.141 & 0.086 & 0.297 \\
2454221.71415 &  -6.716 & 0.098 & 0.339 \\
2454224.70505 &  69.756 & 0.106 & 0.364 \\
2454249.66233 & -19.993 & 0.100 & 0.343 \\
2454251.67471 &  65.708 & 0.101 & 0.349 \\
2454255.67428 & -18.045 & 0.145 & 0.500 \\
\enddata
\end{deluxetable}

\begin{deluxetable}{rrr}
\tabletypesize{\scriptsize}
\tablecaption{Orbital Properties - TYC 272\label{tyc0272_dynprop}}
\tablewidth{0pt}
\tablehead{
\colhead{Parameter} & \colhead{Value} & \colhead{1-$\sigma$ Uncertainty}
}
\startdata
Period (days) & 5.7282 & 0.0003\\
K ($km s^{-1}$) & 54.584 & 0.294 \\
$e$ & 0.001 & $^{+0.009} _{-0.001}$ \\[1ex]
$\omega$ & 5.548 & 0.006 \\
$T_{p}$ &  2454102.9539 & 0.0074 \\
Sys. vel. $\gamma_{0}$ (km s$^{-1}$) & 34.430 & 0.231 \\
$\chi^2/{\rm dof}$ & 2.62 & - \\
\enddata
\end{deluxetable}

\begin{deluxetable}{rrr}
\tabletypesize{\scriptsize}
\tablecaption{Stellar Properties - TYC 272 A\label{tyc0272_starprop}}
\tablewidth{0pt}
\tablehead{
\colhead{Parameter} & \colhead{Value} & \colhead{1-$\sigma$ Uncertainty}
}
\startdata
$\alpha$ (J2000)\tablenotemark{a} & 176.54255245~ (deg)  & 11:46:10.21~ (HH:MM:SS)  \\
$\delta$ (J2000)\tablenotemark{a} & 1.68753726~ (deg)  & +01:41:15.13~ (DD:MM:SS)  \\
$FUV$\tablenotemark{b} & 21.443 & 0.113 \\
$NUV$\tablenotemark{b} & 15.398 & 0.0042 \\
$B$   & 10.945 & 0.035 \\
$V$   & 10.266 & 0.018 \\
$R_c$ &  9.849 & 0.022 \\
$I_C$ &  9.460 & 0.035 \\
$u^\prime$ & 12.055 & 0.100 \\
$g^\prime$ & 10.547 & 0.010 \\
$r^\prime$ & 10.045 & 0.025 \\
$i^\prime$ &  9.880 & 0.020 \\
$z^\prime$ &  9.800 & 0.026 \\
$J_{\rm 2MASS}$\tablenotemark{c} &  8.893 & 0.021 \\
$H_{\rm 2MASS}$\tablenotemark{c} &  8.520 & 0.036 \\
$Ks_{\rm 2MASS}$\tablenotemark{c} &  8.443 & 0.023 \\
$\mu_{\alpha} \left(\rm{mas ~ yr ^{-1} \cos\left(\delta\right)}\right)$\tablenotemark{a} & 146.9 & 1.6 \\
$\mu_{\delta} \left(\rm{mas ~ yr^{-1}}\right)$\tablenotemark{a} & -138.3 & 1.6 \\
$\rm{RPM_{J}}$ & 5.417 & - \\
\enddata
\tablenotetext{a}{Tycho-2 Catalog \citep{hog2000}}
\tablenotetext{b}{GALEX \citep{mar2005}}
\tablenotetext{c}{2MASS \citep{skr2006} Point Source Catalog}
\end{deluxetable}

\begin{deluxetable}{rrr}
\tabletypesize{\scriptsize}
\tablecaption{Light Curve Properties - TYC 272\label{0272lcparams}}
\tablewidth{0pt}
\tablehead{
\colhead{Parameter} & \colhead{Value} & \colhead{1-$\sigma$ Uncertainty}
}
\startdata
$J ~$(Surface brightness ratio at disk center) & 0.14 & 0.01 \\[1ex]
$\frac{R_{1} + R_{2}}{a}$ & 0.096 & 0.004 \\[1ex]
$\frac{R_{2}}{R_{1}}$ & 0.60 & 0.12 \\[1ex]
$i~$(degrees) & 86.6 & 0.3 \\[1ex]
$e\cos{\omega}$ & 0.0004 & 0.0009 \\
$e\sin{\omega}$ & -0.002 & 0.001 \\
Period (days) & 5.72840 & 0.00001 \\
$T_{0} ~$(HJD) & 2454534.7020 & 0.0007 \\[1ex]
$\frac{R_{1}}{a}$ & 0.060 & 0.002 \\[1ex]
$\frac{R_{2}}{a}$ & 0.036 & 0.006 \\[1ex]
$\frac{L_{2}}{L_{1}} ~ $(400-700 nm band) & 0.048 & 0.026 \\
$e$ & 0.002 & 0.0008 \\
$\omega ~$(degrees) & 284 & 101 \\
\enddata
\end{deluxetable}

\begin{deluxetable}{rrrr}
\tabletypesize{\scriptsize}
\tablecaption{MPP RV Observations - TYC 1422\tablenotemark{a}\label{1422rvs}}
\tablewidth{0pt}
\tablehead{
\colhead{HJD$_{\rm{UTC}}$} & \colhead{RV} & \colhead{$\sigma_{RV}$ (formal)} & \colhead{$\sigma_{RV}$ (scaled)}\\
\colhead{~} & \colhead{(km s$^{-1}$)} & \colhead{(km s$^{-1}$)} & \colhead{(km s$^{-1}$)}
}
\startdata
2454101.86376 &  10.848 & 0.090 & 0.276 \\
2454102.03097 & -12.295 & 0.128 & 0.393 \\
2454105.97331 &  67.537 & 0.109 & 0.335 \\
2454128.81685 &  49.394 & 0.099 & 0.304 \\
2454136.77639 &  62.745 & 0.092 & 0.282 \\
2454136.81223 &  53.464 & 0.093 & 0.286 \\
2454164.75574 & -11.161 & 0.106 & 0.325 \\
2454165.75527 &  33.872 & 0.090 & 0.276 \\
2454165.79120 &  33.716 & 0.091 & 0.279 \\
2454186.65694 &  52.251 & 0.091 & 0.279 \\
2454191.72814 &  53.452 & 0.099 & 0.304 \\
2454194.74076 &  13.140 & 0.090 & 0.276 \\
2454195.73280 & -11.254 & 0.090 & 0.276 \\
2454221.62566 &  34.133 & 0.094 & 0.289 \\
2454224.61948 &  33.477 & 0.094 & 0.289 \\
2454254.63340 &  59.181 & 0.129 & 0.396 \\
\enddata
\tablenotetext{a}{These data are dominated by flux contamination and are not used in the final RV analysis.}
\end{deluxetable}

\begin{deluxetable}{rrr}
\tabletypesize{\scriptsize}
\tablecaption{Stellar Properties - TYC 1422 System\label{tyc1422_starprop}}
\tablewidth{0pt}
\tablehead{
\colhead{Parameter} & \colhead{Value} & \colhead{1-$\sigma$ Uncertainty}
}
\startdata
$\alpha$ (J2000)\tablenotemark{a} & 152.41204974~ (deg)  & 10:09:38.89~ (HH:MM:SS)  \\
$\delta$ (J2000)\tablenotemark{a} & 17.58113689~ (deg)  & +17:34:52.09~ (DD:MM:SS)  \\
$g^\prime$ & 10.115 & 0.010 \\
$r^\prime$ & 9.797 & 0.008 \\
$i^\prime$ & 9.726 & 0.013 \\
$J_{\rm 2MASS}$\tablenotemark{b} & 8.875 & 0.026 \\
$H_{\rm 2MASS}$\tablenotemark{b} & 8.641 & 0.028 \\
$Ks_{\rm 2MASS}$\tablenotemark{b} & 8.563 & 0.023 \\
$\mu_{\alpha} \left(\rm{mas ~ yr ^{-1} \cos\left(\delta\right)}\right)$\tablenotemark{a} & -30.6 & 0.9 \\
$\mu_{\delta} \left(\rm{mas ~ yr^{-1}}\right)$\tablenotemark{a} & -5.6 & 1.0 \\
\enddata
\tablenotetext{a}{Tycho-2 Catalog \citep{hog2000}}
\tablenotetext{b}{2MASS \citep{skr2006} Point Source Catalog}
\end{deluxetable}

\begin{deluxetable}{rrrrrrr}
\tabletypesize{\scriptsize}
\tablecaption{SMARTS RV Observations - TYC 1422\label{1422smartsrvs}}
\tablewidth{0pt}
\tablehead{
\colhead{~} & \multicolumn{2}{c}{Component 1} & \multicolumn{2}{c}{Component 2} & \multicolumn{2}{c}{Component 3} \\
\colhead{HJD$_{\rm{UTC}}$} & \colhead{RV} & \colhead{$\sigma_{RV}$} & \colhead{RV} & \colhead{$\sigma_{RV}$} & \colhead{RV} & \colhead{$\sigma_{RV}$} \\
\colhead{~} & \colhead{(km s$^{-1}$)} & \colhead{(km s$^{-1}$)} & \colhead{(km s$^{-1}$)} & \colhead{(km s$^{-1}$)} & \colhead{(km s$^{-1}$)} & \colhead{(km s$^{-1}$)}
}
\startdata
2455236.67830 & 43.107 & 0.298 & -18.420 & 0.602 & 132.754 & 1.650 \\
2455241.64797 & 41.238 & 0.450 & 67.401  & 0.601 & 9.110 & 2.507 \\
2455254.64277 & 31.593 & 10.625 & 31.593  & 10.625 & 72.102 & 1.688 \\
2455270.64841 & 40.024 & 0.488 & 90.128  & 0.490 & -12.069 & 1.648 \\
2455278.59449 & 38.709 & 0.901 & 78.638  & 0.655 & 0.463 & 1.490 \\
2455284.60109 & 39.142 & 0.304 & 87.278  & 0.607 & -8.814 & 1.602 \\
2455293.53225 & 38.567 & 0.381 & -10.746 & 1.071 & 108.512 & 2.509 \\
\enddata
\end{deluxetable}

\begin{deluxetable}{rrrrr}
\tabletypesize{\scriptsize}
\tablecaption{Orbital Properties - TYC 1422\label{tyc1422_dynprop}}
\tablewidth{0pt}
\tablehead{
\colhead{~} & \multicolumn{2}{c}{Component 2} & \multicolumn{2}{c}{Component 3} \\
\colhead{Parameter} & \colhead{Value} & \colhead{1-$\sigma$ Uncertainty} & \colhead{Value} & \colhead{1-$\sigma$ Uncertainty}
}
\startdata
Period (days) & 6.2005 & 0.002 & 6.2005 & 0.002 \\
K ($\rm{km ~ s^{-1}}$) & 65.366 & 0.516 & 84.046 & 1.297 \\
$e\cos{\omega}$ & -0.1494 & 0.0123 & 0.1494 & 0.0123 \\
$e\sin{\omega}$ & 0.0576 & 0.0170 & -0.0576 & 0.0170 \\
$T_{p}$ & 2455236.8133 & 0.0554 & 2455236.8133 & 0.0554 \\
Sys. vel. $\gamma_{0}$ (km s$^{-1}$) & 45.026 & 0.356 & 45.026 & 0.356 \\
Linear Trend $d$ ($\rm{km ~ s^{-1} ~ day^{-1}}$) & -0.048 & 0.018 & -0.048 & 0.018 \\
$\chi^2/{\rm dof}$ & 4.135 & - & 4.135 & - \\
\enddata
\end{deluxetable}

\begin{deluxetable}{rrr}
\tabletypesize{\scriptsize}
\tablecaption{Light Curve Properties - TYC 1422\label{1422lcparams}}
\tablewidth{0pt}
\tablehead{
\colhead{Parameter} & \colhead{Value} & \colhead{1-$\sigma$ Uncertainty}
}
\startdata
$J ~$(Surface brightness ratio at disk center) & 0.88 & 0.01 \\[1ex]
$\frac{R_{1} + R_{2}}{a}$ & 0.162 & 0.002 \\[1ex]
$\frac{R_{2}}{R_{1}}$ & 0.43 & 0.03 \\[1ex]
$i~$(degrees) & 87.7 & 0.5 \\[1ex]
$e\cos{\omega}$ & -0.1283 & 0.0003 \\
$e\sin{\omega}$ & 0.092 & 0.011 \\
$L_{3}~$(Contamination from $3^{\rm{rd}}$ Comp.) & 0.25 & 0.08 \\
Period (days) & 6.199450 & 0.000007 \\
$T_{0} ~$(HJD) & 2454194.3663 & 0.0008 \\[1ex]
$\frac{R_{1}}{a}$ & 0.114 & 0.003 \\[1ex]
$\frac{R_{2}}{a}$ & 0.049 & 0.002 \\[1ex]
$\frac{L_{2}}{L_{1}} ~ $(400-700 nm band) & 0.16 & 0.02 \\[1ex]
$e$ & 0.158 & 0.006 \\
$\omega ~$(degrees) & 144 & 3 \\
\enddata
\end{deluxetable}

\clearpage
%figures here

\end{document}